\documentclass[12pt]{article}
\usepackage{amsmath,amssymb,amsfonts}
\usepackage[dvips]{graphicx}
\usepackage{epsfig}

\makeatletter \@addtoreset{equation}{section} \makeatother

\begin{document}

\begin{titlepage}

\begin{center}
\hfill Imperial/TP/07/SK/01\\
\hfill SNUTP07-006\\
\hfill arXiv:0712.0090

\vspace{1cm}

{\Large\bf The Geometry of Dyonic Instantons\\
\vskip 0.2cm in 5-dimensional Supergravity}

\vspace{1cm}

\renewcommand{\thefootnote}{\alph{footnote}}

{\large Seok Kim$^{1}$\footnote{{\tt s.kim@imperial.ac.uk}}\ \  and\
\ Sungjay Lee$^{2}$\footnote{{\tt saintlee@phya.snu.ac.kr}}}

\vspace{0.7cm}

$^1$\textit{Theoretical Physics Group, Blackett Laboratory,\\
Imperial College, London SW7 2AZ, U.K.}\\

\vspace{0.4cm}

$^1$\textit{Institute for Mathematical Sciences,\\
Imperial College, London SW7 2PE, U.K.}\\

\vspace{0.4cm}

$^2$\textit{School of Physics and Astronomy, Seoul National
University,\\
Seoul 151-747, Korea.}\\

\end{center}

\vspace{0.5cm}

\begin{abstract}

We systematically construct and study smooth supersymmetric
solutions in 5 dimensional $\mathcal{N}\!=\!1$ Yang-Mills-Einstein
supergravity. Our solution is based on the ADHM construction of
(dyonic) multi-instantons in Yang-Mills theory, which extends to the
gravity-coupled system. In a simple supergravity model obtained from
$\mathcal{N}\!=\!2$ theory, our solutions are regular ring-like
configurations, which can also be interpreted as supertubes. By
studying the $SU(2)$ 2-instanton example in detail, we find that
angular momentum is maximized, with fixed electric charge, for
circular rings. This feature is qualitatively same as that of
supertubes. Related to the existence of this upper bound of angular
momentum, we also check the absence of closed timelike curves for
the circular rings. Finally, in supergravity and gauge theory models
with non-Abelian Chern-Simons terms, we point out that the solution
in the symmetric phase carries electric charge which does not
contribute to the energy. A possible explanation from the dynamics
on the instanton moduli space is briefly discussed.

\end{abstract}

\end{titlepage}

\setcounter{footnote}{0}

\tableofcontents

\section{Introduction}

Remarkable progress has been made recently in our understanding of
the supersymmetric solutions in supergravity theories in various
dimensions. The general consequence of the existence of a Killing
spinor has been analyzed in 5 dimensional minimal supergravity
\cite{gghpr}, and then in gauged and/or matter-coupled supergravity
theories \cite{ga-gu1,gu-re,gu-sa} in 5 dimension. Similar studies
in higher dimensions have also been carried out: just to mention a
few of them, 6-dimensional minimal supergravity \cite{gmr},
11-dimensional supergravity \cite{ga-pa,gr-pa} and type IIB
supergravity \cite{gr-pa}.

The general properties of supersymmetric solutions have proven to be
useful in finding new explicit solutions. For instance, in
5-dimensional supergravity theories mentioned above, new black rings
\cite{eemr1,ga-gu3,be-wa,eemr2,ga-gu2} and $AdS_5$ black holes
\cite{gu-re2,gu-re,klr} are discovered, fully utilizing these
structures. A purpose of this paper is to broaden our understanding
to the 5-dimensional supergravity coupled to the vector multiplets
with non-Abelian gauge groups. Technically, this Yang-Mills-Einstein
supergravity is obtained by a procedure called gauging. The gauging
relevant to this theory is that of a global non-Abelian isometry of
the scalar manifold in the vector multiplet, as we review below.

5-dimensional Yang-Mills-Einstein supergravity should have a large
class of supersymmetric solutions, which we expect from our
knowledge of 5-dimensional supersymmetric Yang-Mills field theory in
flat space. Firstly, it is well-known that there are supersymmetric
instanton particles in the latter theory, which are finite energy
solutions of the self-duality equation for Yang-Mills field strength
in spatial $\mathbb{R}^4$, carrying topological charge which we call
the instanton number. The general solution of this non-linear
partial differential equation with finite topological charge is
known, called the ADHM construction \cite{adhm,csw}, which we shall
review and heavily use in this paper. This construction has a
remarkable property of completely solving the self-duality
differential equation, up to an algebraic constraint on the
parameters appearing in the ansatz of the solution. Even if the
latter constraint is notorious as a general closed-form solution is
not available, all the differential equation is completely solved.

A dyonic version of this instanton particle is also known
\cite{la-to}. This configuration carryies electric charge as well as
topological one. It is an instanton particle in the Coulomb phase of
the theory. Ordinary instantons tend to collapse in this phase,
while nonzero electric charge stabilizes this collapse to a finite
size. This `dyonic instanton' has been studied in various
directions, with its interpretation as supertubes \cite{ma-to,kmpw}
(ending on $D4$ branes) \cite{ba-le,ki-le,town,ceh}. The equations
for supersymmetric solutions can again be solved modulo a set of
algebraic constraints, using the ADHM construction
\cite{la-to,dhkm}.

In this paper, firstly, we present the set of general conditions for
the bosonic supersymmetric solutions in 5-dimensional
Yang-Mills-Einstein supergravity, preserving time-like
supersymmetry. This is a simple generalization of \cite{gu-re,gu-sa}
obtained in Maxwell-Einstein supergravity theories. This condition
also generalizes the equations for the dyonic instanton in the field
theory to the gravity-coupled case. A more general analysis of such
conditions is presented in \cite{be-or}, but we shall explain the
derivation to be self-consistent. Secondly, we show that this set of
equations determining the gauge fields, scalars and the metric can
be `solved' in a way which naturally generalizes the ADHM
construction. Namely, we solve all differential equations leaving a
set of algebraic conditions. The solution that we obtain in this
manner is manifestly regular at the generic point of the instanton
moduli space.

From our solution for the metric, one can easily read-off the ADM
angular momentum of the configuration. In models with `rigid'
limits, in which 5 dimensional gauge theory description of
\cite{ims} would become relevant, one naturally expects that same
result could also have been obtained from the Noether angular
momentum in the field theory, which is an integral of angular
momentum density over spatial $\mathbb{R}^4$. The latter integral
could not be evaluated yet. We show that one of the differential
conditions we solve in this paper can be used to make this Noether
integrand into a surface term, giving the same answer as the above
ADM value.

Having the expression for angular momentum and electric charge at
hand, we investigate the $\mathcal{N}\!=\!1$ truncated model of
$\mathcal{N}\!=\!2$ supergravity with $SU(2)$ gauge group in detail.
We find for 2-instanton configurations that various components of
the angular momentum have upper bounds given by the electric
charges, where the maximum is attained when the configuration
becomes a `round circle' on a 2-plane with $U(1)^2$ symmetry like a
ring. This is a feature which also happens for the supertubes
\cite{bho,bhko,ceh}. Our analysis provides another evidence for the
supertube interpretation of our solutions. We also study the
geometry of this $U(1)^2$ symmetric configuration in detail, where
the radius of the `ring' is one of the free parameters of the
solution. In particular, we show that this geometry has no closed
timelike curves (CTC). This should be naturally related to the above
fact that the angular momentum has an upper bound, since it is
over-rotation which usually causes the naked CTC to appear. The
general solution we find does not admit such a source for
over-rotation, which leads us to a conjecture that CTC would be
absent in the general solution we found. We do not attempt to check
it in this paper.

The interpretation of our solution becomes subtler, but interesting,
when there is a non-Abelian Chern-Simons term in the theory for
$SU(N)$ gauge group with $N\geq 3$. For example, such gauge theories
have been obtained from M-theory on singular Calabi-Yau 3-folds
\cite{ims}, where the non-Abelian Chern-Simons coupling arises
either classically or by integrating out massive Dirac fermions.
Since our solution is new even in the gauge theory case, we present
our ADHM solution in the context of both supergravity and gauge
theory. The instanton carries electric charges even in the symmetric
phase, namely, with zero asymptotic VEV for adjoint scalars. The
structure of our general solution suggests a natural model for its
moduli space dynamics, on which we only comment briefly in this
paper.

The rest of this paper is organized as follows. In section 2 we
summarize the necessary backgrounds on 5 dimensional supergravity
coupled to vector multiplets. Special geometry, gauging and several
models are explained. In section 3 we analyze the general structure
of supersymmetric solutions in this theory and derive a set of
differential conditions, generalizing the analysis in the
literature. We also systematically construct regular solutions of
these equations using the ADHM construction. The physical charges,
some of which have been unknown, are computed as well. In section 4
we consider examples. We first consider the properties of gauge
theory solitons, especially in the theory with Chern-Simons
coupling. We also consider the $SU(2)$ 2-instantons in detail: we
find various bounds on physical charges, and identify the structure
of regular ring. Section 5 concludes the paper with discussions.
Derivation of our ADHM solution is given in detail in appendix A.
Properties of Killing spinor bilinears are summarized in appendix B.

\section{Special geometry and gauging}

In this section we summarize some aspects of 5 dimensional
$\mathcal{N}\!=\!1$ Maxwell-Einstein supergravity (preserving 8 real
supersymmetries), and explain the gauging of this theory to obtain
the Yang-Mills-Einstein supergravity. We also explain some models of
our interest. including the related supersymmetric
Yang-Mills-Chern-Simons gauge theory.

The 5 dimensional $\mathcal{N}\!=\!1$ supergravity coupled to $n_V$
Abelian vector multiplets contains the following fields: (1) metric
$g_{\mu\nu}$, (2) gravitino $\psi^i_\mu$ ($i=1,2$), (3) a
graviphoton plus $n_V$ vector fields which are put together and
written as $A^I_\mu$ ($I=1,2,\cdots,n_V+1$), (4) $n_V$ gauginos
$\lambda_i^x$ and (5) $n_V$ real scalars $\varphi^x$ ($x=1,2,\cdots,
n_V$). The coupling of gravity to the vector multiplets is
conveniently described by the real special geometry \cite{gst1}. One
introduces $n_V\!+\!1$ real scalars $X^I$ together with the vector
fields $A^I_\mu$. The scalars $X^I$ have one more degree than is
needed to parameterize $n_V$ dimensional moduli space of
$\varphi^x$, which we call $\mathcal{M}_{n_V}$. $X^I$ is constrained
as
\begin{equation}\label{cubic}
  \mathcal{V}(X)\equiv\frac{1}{6}\ C_{IJK}X^IX^JX^K=1\ ,
\end{equation}
where $C_{IJK}$ is a set of parameters of the theory, totally
symmetric in its indices. When we write $X^I(\varphi^x)$, it is
understood that the above constraint is solved by $\varphi^x$. The
above constraint can be written as
\begin{equation}
  X^IX_I=1\ \ \ {\rm where}\ \ \ \ X_I\equiv\frac{1}{6}\
  C_{IJK}X^JX^K\ .
\end{equation}
The bosonic part of the action of this theory is
\begin{equation}\label{action}
  S=\frac{1}{16\pi G}\int\left(\star R-Q_{IJ}F^I\wedge\star F^J-
  Q_{IJ}dX^I\wedge\star dX^J-\frac{1}{6}C_{IJK}A^I\wedge F^J\wedge F^K\right)
\end{equation}
where we use the metric with mostly plus signature 
and
\begin{equation}
  Q_{IJ}\equiv \frac{9}{2}X_IX_J-\frac{1}{2}C_{IJK}X^K
\end{equation}
is the coupling matrix of $U(1)^{n_V+1}$ gauge fields. This matrix
satisfies $Q_{IJ}X^J=\frac{3}{2}X_I$.

In some theories, including one that we consider in this paper, the
constant $C_{IJK}$ satisfies the so-called symmetric space
condition:
\begin{equation}\label{symm}
  C^{IJK}C_{J(LM}C_{NP)K}=\frac{4}{3}\ \delta^I_{(L} C_{MNP)}\ \ \
  \ \ \ (\ C^{IJK}\equiv C_{IJK}\ )\ .
\end{equation}
In this case, the following relations hold:
\begin{equation}
  \mathcal{V}=\frac{9}{2}\ C^{IJK}X_IX_JX_K\ ,\ \
  X^I=\frac{9}{2}\ C^{IJK}X_JX_K\ .
\end{equation}
The properties of symmetric space are not used when we derive the
supersymmetry conditions or our regular solutions in section 3, but
are used to analyze specific examples in section 4.2.

Now we turn to the gauging of the above theory \cite{gst2,mo-za}. To
this end, we explain the global symmetry of the this theory. The
theory has a global $SU(2)_R$ R-symmetry, which rotates $\psi^i_\mu$
and $\lambda^x_{i}$ as doublets. Apart from this, there can be a
symmetry $G$ which leaves the cubic polynomial $\mathcal{V}(X)$ in
(\ref{cubic}) invariant. The infinitesimal $G$-transformation is
given as
\begin{eqnarray}
  &&\delta X^I=M^I_{\ J}X^J\ ,\ \ \delta A^I_\mu=M^I_{\
  J}A^J_\mu\\
  &&M^I_{\ (J}C_{KL)I}=0\ .
\end{eqnarray}
Leaving the polynomial $\mathcal{V}(X)$ invariant, this
transformation becomes a global symmetry of the Lagrangian, and
especially generates an isometry on $\mathcal{M}_{n_V}$ with the
metric
\begin{equation}
  g_{xy}\equiv Q_{IJ}\frac{\partial X^I}{\partial\varphi^x}
  \frac{\partial X^J}{\partial\varphi^y}\ .
\end{equation}
Among the global symmetry group $SU(2)_R\times G$, we want to gauge
a subgroup $K\subset G$ to obtain Yang-Mills-Einstein
supergravity.\footnote{Another possibility which we do not consider
here is the gauging which includes a subgroup of $SU(2)_R$. In this
case one has to introduce a scalar potential.} We summarize some
aspects of this gauging, referring the readers to \cite{mo-za} and
references therein for details. The gauge field $A^I_\mu$ and the
scalar $X^I$, which are both in an $n_V\!+\!1$ dimensional
(generally reducible) representation of $G$, decompose to
\begin{equation}\label{decompose}
(n_V+1)_G\rightarrow {\rm adj}_K\oplus({\rm singlets})_K\oplus ({\rm
other\ non\ singlets})_K
\end{equation}
under $K$, where ${\rm adj}_K$ denotes the adjoint representation.
The last part consists of the non-singlets apart from the adjoint we
picked out. We label the gauge fields belonging to the adjoint
representation as
\begin{equation}
  A^a_\mu\ ,\ \ a=1,\cdots,\ k\equiv{\rm dim}({\rm adj}_K)\ .
\end{equation}
To gauge the theory with group $K$, one should appropriately insert
the $K$-connection $A^a_\mu$ in the action and supersymemtry
transformations to make this symmetry $K$ a local one: covariantize
the derivatives acting on all non-singlet components of the fields
$X^I(\varphi^x)$ and $\lambda^{x}_i$, change the field strength
$F^a=dA^a$ into a non-Abelian one, and change the Chern-Simons term
into a non-Abelian one. This modification of the action containing
adjoint and other non-singlet fields, if any, breaks the modified
supersymmetry transformation in general. If there are no non-singlet
fields in (\ref{decompose}), the only thing one should do to restore
supersymmetry is to add a suitable Yukawa interaction for fermions
without further deforming supersymmetry transformation rule
\cite{gst2,mo-za}. If there exist non-singlet non-adjoint fields,
one has to work harder to restore supersymmetry \cite{gu-za}.

In this paper, we only consider the case in which the decomposition
(\ref{decompose}) consists of one adjoint and arbitrary number of
singlets. We label the $n_V\!+\!1\!-\!k$ singlet fields as
$A^\alpha_\mu$ and $X^\alpha$. The constants $C_{IJK}$ are
constrained from the symmetry $K$ as
\begin{equation}\label{cubic-restriction}
  C_{abc}=cd_{abc}\ ,\ \ C_{\alpha ab}=C_{\alpha}\delta_{ab}\ ,\ \
  C_{\alpha\beta a}=0\ \ \ \ \
  \left(\ d_{abc}\equiv\frac{1}{2}{\rm tr}(T^a\{T^b,T^c\})\ ,\
  {\rm tr}(T^aT^b)=\frac{1}{2}\delta_{ab}
  \ \right)
\end{equation}
where $T^a$'s are the generators of $K$. $C_{\alpha\beta\gamma}$ is
not constrained. Below we present models of this type derived from
string theory.

The gauging of the subgroup $K\subset G$ outlined above can be done
as follows. The isometry of $\mathcal{M}_{n_V}$ is generated by a
set of Killing vectors. The $k$ Killing vectors $K^x_a(\varphi^x)$
are given as
\begin{equation}\label{killing}
  K^x_a(\varphi)=\frac{3}{2}f^{c}_{\ ab}X_c(\varphi)\left(
  g^{xy}\partial_y X^b(\varphi)\frac{}{}\right)=
  -\frac{3}{2}f^{c}_{\ ab}\left(
  g^{xy}\partial_y X_c(\varphi)\frac{}{}\right)X^b(\varphi)\ ,
\end{equation}
where the second expression is equal to the first one since ($f^a_{\
bc}$ is the structure constant of $K$)
\begin{equation}
  M^L_{\ (I}C_{JK)L}=0\ \rightarrow\ \
  f^L_{\ a(I}C_{JK)L}=0\ \rightarrow\ \ f^{c}_{\ ab}X_cX^b=0\ .
\end{equation}
Firstly the derivatives and field strengths have to be
covariantized. Since our main interest in this paper is to analyze
bosonic solutions, here we only record the bosonic part of the
covariantization:
\begin{eqnarray}
  \partial_\mu\varphi^x&\rightarrow&D_\mu\varphi^x=\partial_\mu
  \varphi^x+gK^x_aA^a_\mu\ ,\\
  F^a_{\mu\nu}=\partial_{\mu}A^a_{\nu}-\partial_{\nu}A^a_{\mu}
  &\rightarrow&F^a_{\mu\nu}=
  \partial_{\mu}A^a_{\nu}-\partial_{\nu}A^a_{\mu}+gf^a_{\ bc}
  A^b_\mu A^c_\nu\ ,
\end{eqnarray}
where $g$ is the coupling constant. Other singlet quantities, like
$F^\alpha_{\mu\nu}$, are unchanged. If $d_{abc}\neq 0$, which is
possible only for $SU(N)$ with $N\geq 3$ among simple groups, the
Chern-Simons term is covariantized to the non-Abelian one:
\begin{equation}
  d_{abc}A^a\wedge dA^b\wedge dA^c\rightarrow tr_{SU(N)}\left(A\wedge F\wedge
  F+\frac{i}{2}A\wedge A\wedge A\wedge F-\frac{1}{10}A\wedge A\wedge
  A\wedge A\wedge A\right)\ ,
\end{equation}
where $F=dA-iA\wedge A$ and $A=A^aT^a$.

Actually, the isometry of our model with (\ref{cubic-restriction})
has a simpler realization as follows. The Killing vector $K^x_a$
transforms $X^I$ as
\begin{equation}
  \delta_a X^I=f^{I}_{\ aJ}X^J=\left\{\begin{array}{ll}
  f^b_{\ ac}X^c&({\rm if}\ I=b)\\0&({\rm if}\ I=\alpha)\end{array}\right.
\end{equation}
which is basically the reason why we required $f^c_{\ a(b}C_{cd)I}=0$ for
the polynomial $\mathcal{V}(X)$ to be invariant under $K$. This
(rather obvious) statement can also be checked directly from the above
definition of $K^x_a$.\footnote{From the definition of Killing vector and
special geometry, one finds
\begin{equation}\label{covariant der}
  \delta_a X^I=\partial_x X^I K^x_a=-\frac{3}{2}f^J_{\ aK}X^K
  \left(g^{xy}\partial_x X^I\partial_y X_J\right)=f^J_{\ aK}X^K
  \left(\delta^I_J-X^IX_J\right)=f^J_{\ aK}X^K
\end{equation}
where structure constants other than $f^c_{\ ab}$ are all zero, and we used
$f^J_{\ IK}X_JX^K=0$. From this we confirm $D_\mu X^I=\partial_\mu X^I+
f^I_{\ bJ}A^b_\mu X^J$ is given as (\ref{covariant der}). Similarly,
one finds  $\delta_a X_I=-f^J_{\ aI}X_J$.}
Therefore one finds
\begin{equation}
  D_\mu X^I=\partial_\mu X^I+gA^a_\mu\delta_a X^I=\left\{
  \begin{array}{ll}\partial_\mu X^a+gf^{a}_{\ bc}A^b_\mu X^c
  &(I=a)\\
  \partial_\mu X^\alpha&(I=\alpha)\end{array}\right.\ ,
\end{equation}
and similarly $D_{\mu}X_a=\partial_\mu X_a+gf^c_{\ ab}A^b_\mu X_c$.

We will sometimes consider the above supergravity together with a
related 5 dimensional Yang-Mills gauge theory model presented in
\cite{ims}. Firstly, we normalize the gauge fields and scalars in
the adjoint representation $(A^a_\mu,X^a)$ such that the covariant
derivatives do not contain the coupling $g$. We define
\begin{equation}
  (A^a_\mu,X^a)_{SYM}=g(A^a_\mu,X^a)_{SUGRA}\ .
\end{equation}
We write $\phi^a\equiv(X^a)_{SYM}$. The first limit we consider is
the one in which the scalars $\phi^a$ and the gauge fields $A^a_\mu$
are `small'. Let us write $\phi^a\sim M$ and
$\frac{\partial_\mu\phi}{\phi}\sim M$, where $M$ is the scale of the
gauge theory, or more specifically of the classical solutions, which
we are interested in. Taking $M\ll g$, we can regard the singlet
scalars $X^\alpha$ as constants (of $\sim\mathcal{O}(1)$). The
metric $g_{\mu\nu}$ can also be taken to be approximately constant
($\approx\eta_{\mu\nu}$), while other singlet fields like
$F^\alpha_{\mu\nu}$ are set to be nearly zero. One also finds
\begin{equation}
  Q_{ab}\approx-\frac{1}{2}C_{abI}(X^I)_{SUGRA}=\frac{1}{2}
  (-C_\alpha X^\alpha)\delta_{ab}-\frac{c}{2g}d_{abc}\phi^c\ .
\end{equation}
If $C_\alpha X^\alpha<0$, one introduces the following Yang-Mills
and Chern-Simons coupling `constants'
\begin{equation}\label{YM coupling}
  \frac{1}{g_{YM}^2}=\frac{(-C_\alpha X^\alpha)}{16\pi Gg^2}\ ,\ \
  c_{YM}=-\frac{c}{16\pi Gg^3}\ .
\end{equation}
The bosonic part of the resulting gauge theory action is given as
\begin{equation}
  S=\int d^5 x\left[-\left(\frac{1}{g_{YM}^2}\delta_{ab}+
  c_{YM}d_{abc}\phi^c\right)\left(\frac{1}{4}F^a_{\mu\nu}F^{b\mu\nu}
  +\frac{1}{2}D_\mu\phi^aD^\mu\phi^b\right)\right]+S_{CS}
\end{equation}
where
\begin{equation}
  S_{CS}=+\frac{c_{YM}}{6}\int{\rm tr}\left(A\wedge F\wedge F+\frac{i}{2}
  A\wedge A\wedge A\wedge F-\frac{1}{10}A\wedge A\wedge A\wedge A\wedge
  A\right)\ .
\end{equation}
This theory can be obtained from the prepotential
$\mathcal{F}(\phi)=\frac{1}{2g_{YM}^2}\phi^a\phi^a+\frac{c_{YM}}{6}d_{abc}
\phi^a\phi^b\phi^c$ \cite{ims}, analogous to $\mathcal{V}(X)$
appearing in (\ref{cubic}). Demanding that the exponential of the
Chern-Simons term be invariant under large gauge transformations,
$c_{YM}$ should be $\frac{1}{(2\pi)^2}$ times an integer \cite{ims},
which can be checked from $\frac{1}{n!}\int_{\mathbb{R}^{2n}}{\rm
tr}(F\wedge\cdots\wedge F)\in(2\pi)^n\mathbb{Z}$.


We close this section by explaining some supergravity models that
will be considered in this paper.

We shall consider in some detail a supergravity model obtained by an
$\mathcal{N}=1$ truncation of the $\mathcal{N}=2$ supergravity. The
latter can be obtained as a low-energy theory of type II string
theory on $K3\times S^1$ or its various U-duals like heterotic
string theory on $T^5$. We start from the $\mathcal{N}=2$
supergravity coupled to $n$ vector multiplets. The latter vector
multiplets contain $n$ gauge fields, $2n$ symplectic Majorana
fermions and $5n$ real scalars. Especially, the scalar manifold is
given as
\begin{equation}\label{N=2 moduli}
  SO(1,1)\times\frac{SO(5,n)}{SO(5)\times SO(n)}
\end{equation}
up to discrete quatients, where the first factor comes from the
dilaton in the $\mathcal{N}\!=\!2$ gravity multiplet. We consider an
$\mathcal{N}=1$ truncation of this $\mathcal{N}=2$ theory, keeping
only the $\mathcal{N}=1$ gravity and vector multiplets while setting
the hypermultiplets and gravitino multiplet to zero.
The scalars $\varphi^x$ in the truncated model live
on the $n_V=n+1$ dimensional space
\begin{equation}\label{N=1 moduli}
 \mathcal{M}_{n_V}=SO(1,1)\times
 \frac{SO(1,\ n_V\!-\!1)}{SO(n_V\!-\!1)}\ ,
\end{equation}
whose special geometry is determined by the polynomial
\begin{equation}\label{polynomial}
  \mathcal{V}(X)=\frac{1}{2}X^1\left(\eta_{ab}X^aX^b\right)\ ,
\end{equation}
where $a,b=2,3,\cdots,n_V+1$ and $\eta_{ab}=diag(+,-,-,\cdots,-)$.
One can easily show that this set of cubic coefficients,
$C_{1ab}=\eta_{ab}$ and others zero, satisfies the symmetric space
condition (\ref{symm}). There is an obvious global symmetry
$SO(1,n_V-1)$ on $\mathcal{M}_{n_V}$. The group $K$ we would like to
gauge is in its compact subgroup, $K\subset SO(n_V\!-\!1)$.

For the above string theory compactification, the massless scalar
moduli is generically given by (\ref{N=2 moduli}) or (\ref{N=1
moduli}) with $n=21$. Near certain points of the moduli space,
namely the fixed points of the isometry $K$, the $U(1)^{21}$ gauge
symmetry enhances to non-Abelian symmetry which technically is
realized as the supergravity gauging. A simple example, among many
others, is $SO(32)\times U(1)^5$ or $E_8\times E_8\times U(1)^5$
where the non-Abelian factors may be regarded as being inherited
from 10 dimensional heterotic gauge symmetry for cetain values of
the moduli. At the level of supergravity, the gauging of the theory
described by (\ref{polynomial}) with respect to \textit{any} Lie
group $K$ can be done by first enlarging the scalar manifolds as
\begin{equation}
  SO(1,1)\times\frac{SO(1,n)}{SO(n)}\ \rightarrow\ \
  SO(1,1)\times\frac{SO(1,k-r+n)}{SO(k-r+n)}\ ,
\end{equation}
where $k$ and $r$ are the dimension and rank of $K$, respectively.
The cubic polynomial is (\ref{polynomial}) with
$a,b=2,\cdots,k\!-\!r\!+\!n\!+\!2\equiv n_V\!+\!1$. The matrix
$\eta_{ab}$ becomes $-\delta_{ab}\propto{\rm tr}(T_aT_b)$ in the
$k$-dimensional subspace with negative signature, proportional to
the quadratic Casimir of any group $K$ of dimension $k$.
$\mathcal{V}(X)$ is therefore invariant under the action of $K$,
which can can be gauged. Under $K$, the $n_V\!+\!1=n+k-r+2$
dimensional representation decomposes as
$(adj)_k\oplus(n\!-\!r\!+\!2\ {\rm singlets})$, which is the class
of theory we discussed. For instance, taking $k=496$ and $r=16$, one
can gauge either subgroup $SO(32)$ or $E_8\times E_8$ of $SO(496)$.

Another interesting example is obtained from M-theory on $K3$-fibred
Calabi-Yau 3-folds \cite{mo-za}. In order to correctly gauge these
models, one has to take care of the 1-loop effect of massive Dirac
fermions in the adjoint representation of $K$, renormalizing the
Chern-Simons coupling $C_{IJK}$. This model is not treated in this
paper. We just mention that there is no such renormalization in the
above $\mathcal{N}=2$ theory due to the underlying $16$
supersymmetry.

\section{Supersymmetric regular solutions}

\subsection{General properties of supersymmetric solutions}

In this section we investigate the general supersymmetric solutions
in the Yang-Mills-Einstein supergravity explained in the previous
section. The strategy is closely related to the ones in, e.g.,
\cite{gghpr,gu-re,gu-sa}. Conventions on geometry and spinors
follows \cite{gghpr}. Especially we use mostly negative metric
$\eta_{\mu\nu}=(+----)$ \textit{only in this subsection and Appendix
B}, to parallel our results with the similar ones in
\cite{gu-re,gu-sa}. To go to the latter convention, changing sign in
front of the Einstein-Hilbert term and the scalar kinetic term would
suffice in the bosonic action (\ref{action}).

We start by assuming the existence of a Killing spinor $\epsilon^i$
($i=1,2$) in a purely bosonic background, satisfying the following
equations coming from the supersymmetry transformations of gravitino
and gaugino:
\begin{equation}\label{gravitino}
  0=\delta\psi^i_\mu=\left(\nabla_\mu+\frac{1}{8}X_I
  (\gamma_{\mu}^{\nu\rho}-4\delta_\mu^{\ \nu}\gamma^\rho)
  F^I_{\nu\rho}\right)\epsilon^i
\end{equation}
and
\begin{equation}\label{gaugino}
  0=\delta\lambda^i_x=\left(\frac{1}{4}Q_{IJ}\gamma^{\mu\nu}
  F^J_{\mu\nu}+\frac{3}{4}\gamma^\mu D_\mu X_I\right)\epsilon^i\
  \frac{\partial X^I}{\partial\varphi^x}\ .
\end{equation}
Here $\nabla_\mu$ denotes the spacetime-covariant derivative, while
$D_\mu$ (acting on $X_I$) is used to emphasize that it is
$K$-covariantized. Its action on $X_I$ is given as
\begin{equation}
  D_\mu X_I=\partial_\mu X_I+f^K_{\ IJ}A^J_\mu X_K\ \ \
  ({\rm where}\ f^{\alpha}_{\ \ast\ast}=f^{\ast}_{\ \ast\alpha}=0)\ ,
\end{equation}
while its action on $F^I_{\mu\nu}$ should also include Christoffel
connection in curved spaces. Using the property $X_I\partial_x
X^I=0$ of special geometry, the gaugino equation (\ref{gaugino}) can
be written as \cite{gu-re,gu-sa}
\begin{equation}
  0=\left(\left(\frac{1}{4}Q_{IJ}-\frac{3}{8}X_IX_J\right)F^J_{\mu\nu}
  \gamma^{\mu\nu}+\frac{3}{4}\gamma^\mu D_\mu
  X_I\right)\epsilon^a\ .
\end{equation}
A bosonic configuration solving the above equation, should
additionally satisfy the equation of motion for the gauge fields
(including the Gauss' law) to be a solution. This equation is
\begin{equation}\label{gauss}
  D(Q_{IJ}\star F^J)=-\frac{1}{4}C_{IJK}F^J\wedge F^K+
  Q_{JK}f^{J}_{\ IL}X^L(\star DX^K)\ .
\end{equation}
Assuming this equation, other equations of motion will turn out to
be guaranteed from the integrability of Killing spinor equation, in
the case we consider (in which timelike supersymmetry is preserved,
to be explained below).

Having a solution of the equations (\ref{gravitino}) and
(\ref{gaugino}), it is helpful to study the various spinor bilinears
following, for instance, \cite{gghpr,gu-re,gu-sa}:
\begin{eqnarray}
  \bar\epsilon^i\epsilon^j&=&f\epsilon^{ij}\\
  \bar\epsilon^i\gamma_\mu\epsilon^j&=&V_\mu\epsilon^{ij}\\
  \bar\epsilon^i\gamma_{\mu\nu}\epsilon^j&=&\Phi^{ij}_{\mu\nu}
  \ \ \ (i\leftrightarrow j\ {\rm symmetric})\ ,\\
  {\rm real\ 2\ forms}\ J^a_{\mu\nu}&:&
  \Phi^{11}=J^1\!+\!iJ^2\ ,\ \ \Phi^{22}=J^1\!-\!iJ^2\ ,\ \
  \Phi^{12}\!=\!-iJ^3\ .
\end{eqnarray}
They satisfy a set of algebraic relations due to Fierz identity, and
differential conditions obtained by using the Killing spinor
equation. The structure of these conditions are similar to the ones
presented in \cite{gghpr,gu-re,gu-sa} and are summarized in appendix
B. Firstly, all algebraic conditions and differential condition
obtained from gravitino equation (\ref{gravitino}) are same as the
results \cite{gu-sa} for the Maxwell-Einstein theory. There are
minor difference in conditions obtained from the gaugino equation
(\ref{gaugino}) and the equation of motion (\ref{gauss}), modified
by the gauging.

Equations (\ref{diff 2}) shows that $V$ is a Killing vector. From
(\ref{alg 1}), it may be either timelike or null. In this paper we
consider the timelike case, which is what we meant by timelike
supersymmetry. Introducing coordinates $(t,x^m)$
($m\!=\!1,\cdots,4$) such that $V=\frac{\partial}{\partial t}$, the
metric can be written as
\begin{equation}\label{metric}
  -ds^2=-f^2(dt+\omega)^2+f^{-1}h_{mn}dx^mdx^n
\end{equation}
where $f$, $\omega$ and $h_{mn}$ are independent of $t$. $h_{mn}$ is
a metric on 4 dimensional base space, which we call $\mathcal{B}$.
Following \cite{gu-sa}, we set $e^0=f(dt+\omega)$, choose the volume
form $(vol)_4$ of $\mathcal{B}$ and take $e^0\wedge (vol)_4$ to be
the 5 dimensional volume form. With $(vol)_4$, we can decompose
$d\omega$ as
\begin{equation}
  fd\omega=G^+ +G^-\ ,
\end{equation}
namely into self-dual and anti-self-dual 2-forms on $\mathcal{B}$,
again following the above references. One can see from (\ref{alg 3})
and (\ref{alg 4}) that $J^i$ can all be regarded as anti-self-dual
2-forms on $\mathcal{B}$, while from (\ref{alg 5}) and (\ref{diff
4}) that they provide an integrable hyper-K\"{a}hler structure on
$\mathcal{B}$ \cite{gghpr}.

Now we turn to the gauge fields. $A^I$ can be written as
$A^I=A^I_0e^0+\mathcal{A}^I$ where $\mathcal{A}^I$ is a 1-form on
$\mathcal{B}$. We choose the gauge $A^I_0=X^I$, which is not
essential but convenient:
\begin{equation}\label{potential}
  A^I=X^Ie^0+\mathcal{A}^I\ .
\end{equation}
Using (\ref{diff 7}), one can follow \cite{gu-sa} and write
\begin{equation}\label{gauge-expression}
  F^I=-f^{-1}e^0\wedge D(fX^I)+\Psi^I+\Theta^I+X^IG^+
\end{equation}
where $\Theta^I$ and $\Psi^I$ are self-dual and anti-self-dual on
$\mathcal{B}$, respectively. Inserting this expression into
(\ref{diff 1}) and (\ref{diff 3}), one obtains
\begin{equation}\label{magnetic-constraint}
  X_I\Psi^I=G^-\ ,\ \ X_I\Theta^I=-\frac{2}{3}G^+\ .
\end{equation}
However, since (\ref{diff 6}) requires $\Psi^I$ to be proportional
to $X^I$, one finds
\begin{equation}
  \Psi^I=X^IG^-\ .
\end{equation}
Inserting this back to (\ref{gauge-expression}), one obtains
\begin{equation}\label{gauge-final}
  F^I=D(X^Ie^0)+\Theta^I\ ,
\end{equation}
where $\Theta^I$ is related to $G^+$ as (\ref{magnetic-constraint}).
Since this field strength is related to the potential
(\ref{potential}) as $F^I=dA^I+\frac{1}{2}f^I_{\ JK}A^J\wedge A^K$,
which is
\begin{equation}
  dA^I+\frac{1}{2}f^I_{\ JK}A^J\wedge A^K=D(X^Ie^0)+\left(d\mathcal{A}^I
  +\frac{1}{2}f^I_{\ JK}\mathcal{A}^J\wedge\mathcal{A}^K\right)\ ,
\end{equation}
one concludes that the self-dual component $\Theta^I$ is given by the
1-form $\mathcal{A}^I$ on $\mathcal{B}$ as
\begin{equation}\label{field-strength}
  \Theta^I=d\mathcal{A}^I+\frac{1}{2}f^I_{\ JK}\mathcal{A}^J\wedge
  \mathcal{A}^K\ ,
\end{equation}
which is exactly the Yang-Mills field strength of $\mathcal{A}^I$ on
the space $\mathcal{B}$. The set of constraints on $\Theta^I$ is
(\ref{field-strength}), self-duality on $\mathcal{B}$, and
(\ref{magnetic-constraint}).

Following the Maxwell-Einstein supergravity, one can also show that
the above conditions are sufficient to show the Killing spinor
equations. Firstly, imposing the projection
$\gamma^0\epsilon^i=\epsilon^i$, the gaugino equation follows from
(\ref{gauge-final}) and the fact $\Theta^I=\star_4\Theta^I$. The
gravitino equation reduces to
\begin{equation}
  \partial_t \epsilon^i=0\ ,\ \
  \nabla_m(f^{-\frac{1}{2}}\epsilon^i)=0\ .
\end{equation}
As in the Maxwell-Einstein theory, there exist 4 real independent
components solving these equations and
$\gamma^0\epsilon^i=\epsilon^i$ on the hyper-K\"{a}hler space
$\mathcal{B}$.

Apart from the conditions for supersymmetry, one also has to impose
the equation of motion for the gauge fields. After imposing the
supersymmetry conditions, it turns out that the only nontrivial
component of this equation is the Gauss' law:
\begin{equation}
  \mathcal{D}^m\mathcal{D}_m(f^{-1}X_I)=\frac{1}{6}C_{IJK}\star_4
  (\Theta^j\wedge\Theta^K)\ .
\end{equation}
As mentioned above, the supersymmetry conditions and this Gauss' law
guarantee other equations of motion also hold in our timelike case.

To summarize, one obtains the following set of equations to be
solved:
\begin{eqnarray}
  \Theta^I&=&\star_4\Theta^I\ \ \ \
  (\Theta^I=d\mathcal{A}^I+f^I_{\
  JK}\mathcal{A}^J\wedge\mathcal{A}^K)\label{final 1}\\
  \mathcal{D}^m\mathcal{D}_m(f^{-1}X_I)&=&\frac{1}{6}C_{IJK}\star_4
  \left(\Theta^J\wedge\Theta^K\right)\label{final 2}\\
  (1+\ast_4) d\omega&=&-3f^{-1}X_I\Theta^I\label{final 3}
\end{eqnarray}
where $\mathcal{D}_m$ is the covariant derivative on $\mathcal{B}$
with the connection $\mathcal{A}^I$. These equations should be
solved to give the fields $\mathcal{A}^I$, $X^I$, $f$ and $\omega$.
The basic fields are given by (\ref{metric}) and (\ref{potential}).
The above three equations are similar to those in the Maxwell theory
\cite{be-wa,ga-gu2}. There, if one tries to solve them in the order
listed above, they can be regarded as linear equations with source.
The situation is nearly the same here. The first equation is
non-linear to start with. However, the latter two can be solved
linearly, regarding the right hand sides as external source terms
once the previous equations are solved. Even the first non-linear
equation is has been studied in depth, since it is the famous
equation describing self-dual instantons in Yang-Mills theory. In
the next subsection, we present a large class of (semi-)explicit
solutions of this set of equations.

\subsection{ADHM instantons and regular solutions}

From now we assume the base space $\mathbb{R}^4$ and systematically
find a class of configurations solving (\ref{final 1}), (\ref{final
2}), (\ref{final 3}). The self-dual Yang-Mills gauge field
configurations on $\mathbb{R}^4$ can be found by the so-called ADHM
construction \cite{adhm,csw}. We base our analysis on the ADHM
construction to find solutions of the other equations we listed in
the previous subsection.

Before starting the analysis, we would like to clarify the different
normalizations in supergravity and Yang-Mills theory. So far we
naturally normalized the scalars $X^I$ and gauge fields $A^I_\mu$ to
have mass dimension $0$. The gauge coupling $g$ has dimension $1$. A
convenient normalization for the analysis of solitons in gauge
theory is to set this coupling to $1$ by rescaling $X^I_{YM}=g
X^I_{SUGRA}$ and $A^I_{YM}=gA^I_{SUGRA}$, where the prefactor in
front of the kinetic terms of vector multiplet fields becomes
$\frac{1}{16\pi G g^2}$. We assume the latter normalization in this
subsection and Appendix A. In this normalization, scalars satisfy
$\frac{1}{6}C_{IJK}X^IX^JX^K=g^3$. The equations (\ref{final 1}) and
(\ref{final 2}) takes the same form replacing
$X_I\equiv\frac{1}{6}C_{IJK}X^JX^K$ and $\Theta^I$ into the new
ones, while (\ref{final 3}) becomes
\begin{equation}\label{final 3-2}
  (1+\star_4)d\omega=-3g^{-3}(f^{-1}X_I)\Theta^I
\end{equation}
with the new normalization.

As mentioned above, we choose the 4 dimensional base space to be
$\mathbb{R}^4$ with the flat metric $h_{mn}=\delta_{mn}$, even
though there are more general possibilities of base space. With this
choice of the base space, the general solution to the self-dual
field equation (\ref{final 1}) is given by the ADHM construction
which we explain now. We will exclusively consider the case with
$SU(N)$ gauge group in this paper, even if we expect the cases with
$SO(N)$ and $Sp(N)$ gauge groups can be treated in a similar way.
Following \cite{dhkm}, one starts the construction of $SU(N)$
$k$-instantons by writing down an $(N+2k)\times 2k$ matrix
$\Delta_{\dot\alpha}(x)$
\begin{equation}
  \Delta_{\dot\alpha}\equiv a_{\dot\alpha}+b^ax_{\alpha\dot\alpha}
\end{equation}
where
\begin{equation}
  x_{\alpha\dot\alpha}\equiv x^m\sigma^m_{\alpha\dot\alpha}\ ,\ \
  x^m\in \mathbb{R}^4\ ,\sigma^m_{\alpha\dot\alpha}=(1,i\vec\sigma)\ ,
  \bar\sigma^{n\dot\alpha\alpha}=(1,-i\vec\sigma)
\end{equation}
and
\begin{equation}
  a_{\dot\alpha}\equiv
  \left(\begin{array}{c}\omega_{\dot\alpha}\\a^\prime_{\alpha\dot\alpha}
  \end{array}\right)\ ,\ \ b^\alpha\equiv
  \left(\begin{array}{c}{\bf{0}}_{N\times 2k}\\{\bf{1}}_{2k\times 2k}
  \end{array}\right)\ .
\end{equation}
The constant matrices $\omega_{\dot\alpha}$ and
$a^\prime_{\alpha\dot\alpha}\equiv a_n \sigma^n_{\alpha\dot\alpha}$,
$a_n$ are $N\times 2k$, $2k\times 2k$ and $k\times k$ matrices,
respectively, and we suppressed all matrix indices except for the
2-component $SO(4)$ spinor indices $\alpha$ and $\dot\alpha$. We
refer the readers to \cite{dhkm} for more details on notations.

The self-dual field strength $\Theta_{mn}$, or the connection
$\mathcal{A}_m$, is given by an $(N+2k)\times N$ matrix $U(x)$
satisfying the following conditions
\begin{equation}
  \bar\Delta^{\dot\alpha}(x)U(x)=0\ ,\ \ \bar{U}U={\bf 1}_{N\times N}\ .
\end{equation}
The gauge field $\mathcal{A}_m$ is given as
\begin{equation}
  \mathcal{A}_m=i\bar{U}(x)\partial_mU(x)\ ,
\end{equation}
whose field strength is guaranteed to be self-dual if
$\omega_{\dot\alpha}$ and Hermitian matrices $a_n$ satisfy the
following algebraic equation
($\bar\sigma_{mn}\equiv\bar\sigma_{[m}\sigma_{n]}$ and
$\sigma_{mn}\equiv\sigma_{[m}\bar\sigma_{n]}$):
\begin{equation}\label{ADHM constraint}
  \bar\omega^{\dot\alpha}\omega_{\dot\beta}
  (\bar\sigma_{mn})^{\dot\beta}_{\ \dot\alpha}=2(1-\ast_4)[a_m,a_n]\ .
\end{equation}
With (\ref{ADHM constraint}) satisfied, one can show that the
$2k\times 2k$ matrix $\bar\Delta^{\dot\alpha}\Delta_{\dot\beta}$
takes the form
\begin{equation}\label{invertible}
  \bar\Delta^{\dot\alpha}\Delta_{\dot\beta}=F^{-1}(x)
  \delta^{\dot\alpha}_{\dot\alpha}
\end{equation}
with an invertible $k\times k$ Hermitian matrix $F(x)$. The field
strength $\Theta^aT^a$, where $T^a$'s are $SU(N)$ generators with
the normalization in section 2, is given as
\begin{equation}
  \Theta_{mn}\equiv\Theta^a_{mn}T^a=2i\bar{U}
  b^\alpha(\sigma_{mn})_\alpha^{\ \beta}F\bar{b}_{\beta}U\ .
\end{equation}
The general solution to the $k\times k$ matrix equation (\ref{ADHM
constraint}) is not known, but we will say that one `solved' the
equation (\ref{final 1}) in the sense that partial differential
equation is reduced to an algebraic one. The number of unconstrained
real degrees in the matrices are $4Nk$: from the original $4Nk+4k^2$
degrees in $\omega_{\dot\alpha}$ and $a_n$, one subtracts the number
of equations in (\ref{ADHM constraint}), $3k^2$, as well as the
$U(k)$ gauge transformation degree $k^2$ \cite{dhkm}. This actually
is the general self-dual configuration with given topological charge
$k$, deduced from a suitable index theorem.

Having this general solution parameterized by $4Nk$ data, one has to
solve the covariant Laplace equation with sources (\ref{final 2}).
We first consider the scalars in the adjoint representation, $I=a$.
The Laplace equation without source is solved in \cite{dhkm}, see
their Appendix C. In Appendix A.1, we generalize this construction
to the case with sources provided by the non-Abelian Chern-Simons
term. The equation and our solution in matrix notation are
\begin{equation}\label{matrix scalar equation}
  \mathcal{D}^2(f^{-1}X_aT^a)=\frac{c}{24}\left(\Theta_{mn}
  \Theta_{mn}-\frac{1}{N}{\rm tr}(\Theta_{mn}\Theta_{mn}){\bf 1}_N
  \right)
\end{equation}
and
\begin{equation}\label{scalar charged}
  f^{-1}X_aT^a=\bar{U}(x)\mathcal{J}_0U(x)-\frac{c}{24N}
  \partial^2\log(\det F(x))\ {\bf 1}_N
\end{equation}
where the $(N\!+\!2k)\times (N\!+\!2k)$ matrix $\mathcal{J}_0$ is
given as
\begin{equation}
  \mathcal{J}_0=\left(\begin{array}{cc}v_{N\times N}&\\&
  \left(\varphi_{k\times k}-\frac{c}{12}F(x)\right)\otimes{\bf 1}_2
  \end{array}\right)\ .
\end{equation}
We hope using $\varphi$ will not cause confusion with scalars
$\varphi^x$ in section 2. Here the $k\times k$ matrix $\varphi$
should satisfy
\begin{equation}\label{algebraic}
  {\bf L}\varphi\equiv\frac{1}{2}\left\{\bar\omega^{\dot\alpha}
  \omega_{\dot\alpha},\varphi\right\}+\left[a_n,\left[a_n,\varphi\right]
  \right]=\bar\omega^{\dot\alpha}v\omega_{\dot\alpha}-
  \frac{c}{6}{\bf 1}_k
\end{equation}
for (\ref{scalar charged}) to solve (\ref{matrix scalar equation}).
The $N\times N$ matrix $v=v_aT^a$ is the asymptotic value of $X_aT^a$
at infinity. Equation (\ref{algebraic}) is linear in $\varphi$. We will 
present the explicit 2-instanton solutions in the next section. Anyhow, the
differential equation is solved modulo the algebraic
equation (\ref{algebraic}).

Now we turn to the Laplace equation for the singlet scalars with
source terms,
\begin{equation}
  \partial^2(f^{-1}X_\alpha)=\frac{C_\alpha}{6}\ast_4(\Theta^a\wedge
  \Theta^a)=\frac{C_\alpha}{6}\ {\rm tr}\left(\Theta_{mn}\Theta_{mn}
  \right)\ .
\end{equation}
It can be solved using the Osborn's
formula \cite{osborn} for the topological charge density:
\begin{equation}
  {\rm tr}_N(\Theta_{mn}\Theta_{mn})=(\partial^2)^2
  \log\left(\det F_{k\times k}(x)\frac{}{}\right)\ .
\end{equation}
From this one obtains
\begin{equation}\label{scalar neutral}
  f^{-1}X_\alpha=+\frac{1}{6}C_\alpha \partial^2\log
  \left(\det F(x)\frac{}{}\right)+h_\alpha\ ,
\end{equation}
where $h_\alpha$ are constants. One might have inserted any harmonic
function $H_\alpha(x)$ on $\mathbb{R}^4$ instead of $h_\alpha$,
which is a homogeneous solution of this equation. However, in
foresight, we do not insert any nontrivial homogeneous solution,
which in $\mathbb{R}^4$ is associated with singular sources, in
order to obtain regular solutions.

Finally, we turn to the differential equation (\ref{final 3}) for
the 1-form $\omega_m$. In Appendix A.2, we derive the following
solution in general ADHM instanton background:
\begin{equation}\label{omega result}
  \omega_m=-\frac{3i}{g^3}\
  {\rm tr}\left(\mathcal{J}_0\frac{\mathcal{P}\partial_m
  \mathcal{P}-\partial_m\mathcal{P}\mathcal{P}}{2}+
  2[\varphi,a_m]F-\frac{c}{72}\epsilon_{mnpq}
  \partial_{n}F^{-1}F\partial_{p}F^{-1}
  F\partial_qF^{-1}F\right)\ ,
\end{equation}
where $\mathcal{P}(x)\equiv U\bar{U}$. Again, one might add
arbitrary homogeneous solution $\Delta\omega_m$ to the equation
(\ref{final 3}), where $d(\Delta\omega)$ is anti-self-dual. For
nonzero $\Delta\omega_m$ to vanish at asymptotic infinity, it should
also be associated with a singular source since the Maxwell equation
$d^\dag d(\Delta\omega)=0$ is satisfied for anti-self-dual
$d(\Delta\omega)$. For instance, adding
\begin{equation}
  \Delta\omega_m=\frac{j}{r^4}(\delta_m^{[1}x^{2]}+\delta_m^{[3}x^{4]})
\end{equation}
to a spherically symmetric black hole would change the solution into
the BMPV black hole with the self-dual angular momentum
$(J_L)_{12}=(J_L)_{34}\sim j$. Adding it to our solution would
result in closed timelike curves. Anyhow we again do not add such
homogeneous solutions. This completes the construction of our
solution of (\ref{final 1})-(\ref{final 3}).

We emphasize that the solution we obtained is manifestly smooth
`generically': namely all components of the fields ($g_{\mu\nu}$,
$F^I_{\mu\nu}$, $X^I$) are finite and smooth in space-time
coordinates $(t,x^m)$, at a generic point on the instanton moduli
space. This is guaranteed from the construction itself, once the
matrix $F(x)$ introduced in (\ref{invertible}) is invertible. This
assumption is not true on a certain point of the instanton moduli
space. For example, there are parameters which can be identified as
the sizes of instantons. When any of these sizes is taken to zero,
the configuration $\Theta^a_{mn}$ starts to be singular at the
`location' of this small instanton. This singularity propagates to
the other fields at this point. Just to mention one phenomenon, let
us consider the Chern-Simons coupling $C_\alpha A^\alpha\wedge
(F^a\wedge F^a)$ which induces $U(1)$ electric charges of
$A^\alpha_\mu$ to an instanton. As the instanton becomes small, the
souce for $F^\alpha_{\mu\nu}$ becomes point-like, which has an
effect of replacing $h^\alpha$ in (\ref{scalar neutral}) by a
harmonic function sourced by a point charge. Away from such
`singular' points on the instanton moduli space, our configuration
is smooth.

In \cite{gkltt}, regular solutions for the gravitating single
monopoles and instantons in 4- and 5-dimensional (super-)gravity
saturating BPS energy bounds are constructed. Moreover, in the 't
Hooft (dyonic) multi-instanton background, the regular solutions in
10 dimensional heterotic supergravity is obtained in
\cite{stro,etz}. Our solution is a generalization of these works in
the 5 dimensional setting.


We close this section by computing the physical charges
of our solutions.

The $U(1)^{N-1}\subset SU(N)$ electric charge $q$ is given as
(choosing the orientation $dt\wedge dr\wedge
{\rm vol}(S^3)$)
\begin{equation}
  q_a\sim\frac{1}{8\pi G}\int_{S^3}Q_{aI}\star F^I=\frac{3}{16\pi G}
  \int_{S^3}\star_4 \mathcal{D}(f^{-1}X_a)
\end{equation}
where the integral is over the asymptotic 3-sphere. We multiply
$\frac{1}{g^3}$ on the right hand side, which will turn out to be the most
natural normalization. Expanding the integrand, the
electric charge is given by the $N$ diagonal entries of the
following $N\times N$ matrix,
\begin{equation}\label{electric}
  q=-\frac{3}{16\pi Gg^3}\cdot 4\pi^2
  \left(\frac{1}{2}\{v,\ \omega_{\dot\alpha}
  \bar\omega^{\dot\alpha}\}-\omega_{\dot\alpha}\varphi
  \bar\omega^{\dot\alpha}-\frac{ck}{6N}{\bf 1}_N\right)\equiv
  4\pi^2\left(\frac{1}{2}\{\check{v},\ \omega_{\dot\alpha}
  \bar\omega^{\dot\alpha}\}-\omega_{\dot\alpha}\check{\varphi}
  \bar\omega^{\dot\alpha}-\frac{c_{YM}k}{2N}{\bf 1}_N\right)\ ,
\end{equation}
where we introduce new variables
$\check{v}\equiv-\frac{3}{16\pi Gg^3}v$,
$\check\varphi\equiv-\frac{3}{16\pi Gg^3}\varphi$ in foresight. In
section 4.1 we show this is the natural normalization in Yang-Mills
field theory. $c_{YM}$ is already introduced as (\ref{YM coupling}).
The matrix $q$ is traceless if $\varphi$ satisfies
(\ref{algebraic}). This is a simple generalization of the result in
\cite{la-to} to the case $c\neq 0$.

We now compute the ADM angular momentum associated with the
Killing vector $-\xi_{ab}$, where
$\xi_{ab}\equiv x_a\partial_b-x_b\partial_a$:
\begin{equation}
  J_{ab}=-\frac{1}{16\pi G}\int_{S^3}\star\nabla\xi_{ab}
  =-\frac{4\pi^2 k_{ab}}{16\pi Gg^3}
\end{equation}
where $\omega_m\approx\frac{k_{mn}x^n}{g^3r^4}$ as
$r\rightarrow\infty$. Expanding the matrices $F$ and $\mathcal{P}$,
one obtains from (\ref{omega result}) the following:
\begin{eqnarray}
  \omega_m&\approx&\frac{3i}{g^3}\frac{x^n}{r^4}\left\{{\rm tr}_k\left(
  \bar\omega^{\dot\alpha}v\omega_{\dot\beta}-\varphi
  \bar\omega^{\dot\alpha}\omega_{\dot\beta}\right)
  (\bar\sigma_{mn})^{\dot\beta}_{\ \dot\alpha}+4{\rm tr}_k
  \left(\varphi[a_m,a_n]\frac{}{}\right)\right\}\nonumber\\
  &=&\frac{3i}{g^3}\frac{x^n}{r^4}\left\{{\rm tr}_k\left(
  \bar\omega^{\dot\alpha}v\omega_{\dot\beta}\right)
  (\bar\sigma_{mn})^{\dot\beta}_{\ \dot\alpha}+2(1+\ast_4)
  {\rm tr}_k\left(\varphi[a_m,a_n]\frac{}{}\right)\right\}\ ,
\end{eqnarray}
where we used the ADHM constraint (\ref{ADHM constraint}) on the
second line. Therefore, one finally obtains
\begin{equation}\label{ADM angular final}
  J_{mn}=+4\pi^2 i\left\{
  {\rm tr}_k\left(
  \bar\omega^{\dot\alpha}\check{v}\omega_{\dot\beta}\right)
  (\bar\sigma_{mn})^{\dot\beta}_{\ \dot\alpha}+2(1+\ast_4)
  {\rm tr}_k\left(\check{\varphi}[a_m,a_n]\frac{}{}\right)\right\}\ ,
\end{equation}
where again the new variables $\check{v}$ and $\check\varphi$ are
introduced as shown in the previous paragraph, to compare the result
(\ref{ADM angular final}) with the one from
field theory in section 4.1.

The ADM mass is associated with the Killing vector
$\xi=\partial_t$ \footnote{The Killing vector for mass always picks
up a minus sign relative to those for spatial momenta
\cite{gu-re2}.}:
\begin{equation}
  M=+\frac{3\pi\alpha}{4G}
\end{equation}
if $f\approx 1-\frac{\alpha}{r^2}$ as $r\rightarrow\infty$.
With the asymptotic behavior
\begin{equation}
  f^{-1}X_I\approx h_I+\frac{\mu_I}{r^2}\ ,\ \
  f^{-\frac{1}{2}}X^I\approx h^I+\frac{\mu^I}{r^2}\ ,
\end{equation}
where $h^I=X^I(\infty)$, 
one can easily find $\alpha=\mu_Ih^I$. From
(\ref{scalar charged}) and (\ref{scalar neutral}) one finds
%
\begin{equation}\label{ADM mass}
  M=q_a\phi^a(\infty)+\frac{8\pi^2 k}{g_{YM}^2}\ ,
\end{equation}
where $\phi^a(\infty)$ is the expectation value of $\phi^a$($=gX^a_{SUGRA}$)
at infinity, and $g_{YM}^2$ is given by (\ref{YM coupling}). This
saturates the BPS bound given in \cite{la-to}.

\section{Examples and applications}

\subsection{The Yang-Mills(-Chern-Simons) gauge theory}

When the Yang-Mills gauge fields and scalars are taken to be
`small', as explained in section 2, our solution reduces to that of
the gauge theory of \cite{ims}. The dyonic instanton configuration
in the gauge theory without non-Abelian Chern-Simons term has been
first studied in \cite{la-to}. The general ADHM solution in the
presence of the non-Abelian Chern-Simons term has been unknown in
the gauge theory, so we shall take a more detailed look at our new
solution in this context. Another problem in the gauge theory which
has not been answered is the computation of the Noether angular
momentum of the configuration. In the previous section we obtained
the ADM angular momentum, but it seems that the same answer should
be obtained on the gauge theory side as the Noether charge. We also
explain this point in this subsection.

In the Yang-Mills-Chern-Simons theory, the differential conditions
for the supersymmetric solutions are
\begin{eqnarray}
  F_{mn}&=&{\star_4 F\nonumber}_{mn}\\
  \mathcal{D}^2\check\phi_a&=&+\frac{c_{YM}}{4}d_{abc}F^b_{mn}F^{c}_{mn}
\end{eqnarray}
where
\begin{equation}
  \check\phi_a\equiv\frac{\partial\mathcal{F}(\phi)}{\partial\phi^a}=
  \frac{1}{g_{YM}^2}\phi^a+\frac{c_{YM}}{2}d_{abc}
  \phi^b\phi^c\ \ \ \left(\mathcal{F}(\phi)\equiv\frac{1}{2g_{YM}^2}
  \phi^a\phi^a+\frac{c_{YM}}{6}d_{abc}\phi^a\phi^b\phi^c\right)\ .
\end{equation}
Furthermore, even if the metric degree considered in the previous
section is irrelevant in the gauge theory, we would still like to
consider the regular solution of the following diffential equation:
\begin{equation}\label{omega SYM}
  (1+\star_4)(d\alpha)_{mn}=-6{\rm tr}_N(\check\phi F_{mn})\  ,
\end{equation}
where the regular solution for the 1-form $\alpha_m$ can be obtained
as we got $\omega_m$ before. If one takes the scaling of fields
$F_{mn}^{YM}=gF_{mn}^{SUGRA}$ and $\check\phi_a=-\frac{3}{16\pi G
g}X_a^{SUGRA}$ into account, one obtains
\begin{equation}
  \alpha_m=\omega_m(\check{v})=-\frac{3}{16\pi Gg^3}
  \ \omega_{m}(v)\ ,
\end{equation}
with $\omega_m$ given as (\ref{omega cs}). This differential equation
and the solution $\alpha_m$ will still play interesting roles as we
explain below.

Firstly, let us re-consider the electric charge computed in
the previous seciton. The expression (\ref{electric})
is exactly the same as that in \cite{la-to} in the case $c\!=\!0$
(correcting a factor 2 typo there). The quantization
of this electric charge was studied from the moduli space dynamics
of Yang-Mills instantons \cite{la-to}, where the electric charge is
understood as a momentum conjugate to the coordinate on the moduli
space parameterizing the global gauge zero mode. See \cite{pe-za} also.
A potential of the schematic form  $\propto v^2|\omega_{\dot\alpha}|^2$
is generated on the moduli space, which
holds the motion in the moduli space in a finite
$\omega_{\dot\alpha}$ region. Since $\varphi$ is also proportional
to $v$, one finds that the electric charge depens linearly on the
asymptotic value $v^a$ of $\check{\phi}^a=\frac{1}{g_{YM}^2}\phi^a$.

When there is a non-zero Chern-Simons term, $c\!\neq\!0$, the
physics becomes different. In this case one finds that the
configuration carries nonzero electric charge even when
$\phi^a(\infty)=0$ (or $v\!=\!0$), as the second and third terms of
(\ref{electric}) are still nonzero. From the dynamics on the
instanton moduli space, this quantity should also be understood as
the momentum conjugate to the global gauge zero mode. The Lagrangian
should acquire modifications other than the potential to explain
this charge. From (\ref{ADM mass}), the electric charge, or
momentum, does not contribute to the BPS mass if $\phi^a(\infty)=0$.
It is likely that the states with electric charges should provide a
sort of lowest Landau level degeneracy from the viewpoint of moduli
space dynamics, by an addition of external magnetic field on the
moduli space.\footnote{We thank David Tong for pointing it out to
us.}

For simplicity, let us briefly comment on the single instantons in the
unbroken phase ($\phi^a(\infty)=0$) when $c_{YM}\neq 0$. The magnetic
field $\Theta^a_{mn}$ is given by the $SU(N)$ embedding of single
$SU(2)$ 't Hooft solution. In this background, one finds nonzero
scalar and electric field. However, the electric charge
contribution to the energy is zero since $v=0$. One finds
\begin{equation}\label{cs-single-charge}
  \check\varphi=-\frac{c_{YM}}{4\lambda^2}\ \rightarrow\ \
  q=\pi^2 c_{YM}\left(P-\frac{2}{N}{\bf 1}_N\right)\ ,
\end{equation}
where $\lambda$ is the size of the instanton, $P$ is the projector
to the 2 dimensional subspace of the $N$ dimensional space in which
$SU(2)$ 't Hooft solution is embedded. With $\check{v}=0$, since the
potential $\propto\lambda^2\check{v}^2$ confining $\lambda$ is
absent, the nature of the corresponding motion on the moduli space
should be quite different. What we expect from
(\ref{cs-single-charge}) is a motion on the moduli space with
appropriate `magnetic field.' Just for convenience, let us assume
that $\lambda$ is much larger than $c_{YM}g_{YM}^2$, the only scale
of this system. Then we can trust the moduli space metric for single
instantons with $c_{YM}=0$, which is a cone over
$\frac{SU(N)}{U(N\!-\!2)}$ with homogeneous metric on the base. Upon
coupling the system to a suitable 1-form $A\sim c_{YM}\theta$, where
$d\theta$ gives the Kahler 2-form of the space
$\frac{SU(N)}{SU(2)\times U(N\!-\!2)}$, one finds that the rest
particle solution carries an angular momentum of the form
(\ref{cs-single-charge}). More comment is in order in the conclusion
section.

Now we consider the angular momentum of the configuration.
The Noether angular momentum is given by
the following 4 dimensional integral\footnote{Overall minus sign is
inserted since positive energy is given by $\int d^4 x T_{00}$,
while spatial momentum has a relative minus sign in its definition.}:
\begin{equation}\label{Noe angular}
  J_{mn}=-\int d^4 x (x^m T_{0n}-x^nT_{0m})\ ,
\end{equation}
where
\begin{equation}
  T_{0m}=\left(\frac{1}{g_{YM}^2}\delta_{ab}+c_{YM}d_{abc}\phi^c\right)
  F_{0n}^aF_{mn}^b=-2\partial_n{\rm tr}\left(\check{\phi}F_{mn}\right)\ .
\end{equation}
The integral (\ref{Noe angular}) can be written as
\begin{equation}\label{Noe intermediate}
  J_{mn}=-2\int_{S^3}r^3d\Omega^k\left(
  \frac{}{}x^n{\rm tr}\left(\check{\phi}F_{mk}\right)-
  x^m{\rm tr}\left(\check{\phi}F_{nk}\right)\right)+4
  \int_{\mathbb{R}^4} d^4x\ {\rm tr}\left(\check\phi F_{mn}\right)\ ,
\end{equation}
where $d\Omega^k$ is the vector normal to the unit 3-sphere whose
length is the volume element of $S^3$. In \cite{etz}, the second term
is shown to be zero for the 't Hooft multi-instanton background. The
first surface term is easily evaluated to give an expression for
$J_{mn}$ in this case. For general ADHM instanton, the second term
is nonzero and the general expression of $J_{mn}$ has not been
available yet. However, one can also change the second term of
(\ref{Noe intermediate}) into a surface term, using the differential
condition (\ref{omega SYM}):
\begin{equation}
  \int d^4 x {\rm tr}(\check\phi F_{mn})=
  -\frac{1}{6}\int_{S^3}r^3\left(d\Omega^m\alpha_n-
  d\Omega^n\alpha_m+\epsilon_{mnpq}d\Omega^p\alpha_q\frac{}{}\right)
\end{equation}
where we used the following fact for an integral over a region $\Sigma$
in $\mathbb{R}^4$:
\begin{equation}
  \int_{\Sigma}d^4 x\ \partial_a f=\int_{\partial\Sigma}dS^a\ f\ .
\end{equation}
Evaluating the two surface integrals, one finds
\begin{eqnarray}
  -2\int_{S^3}r^3d\Omega^k\left(
  \frac{}{}x^n{\rm tr}\left(\check{\phi}F_{mk}\right)-
  x^m{\rm tr}\left(\check{\phi}F_{nk}\right)\right)&=&
  +4\pi^2 i{\rm tr}\left({\bar\omega}^{\dot\alpha}\check{v}\omega_{\dot\beta}
  ({\bar\sigma}_{mn})^{\dot\beta}_{\ \dot\alpha}\right)\\
  -\frac{2}{3}\int_{S^3}r^3\left(d\Omega^m\alpha_n-d\Omega^n\alpha_m+
  \epsilon_{mnpq}d\Omega^p\alpha_q
  \right)&=&+\frac{2\pi^2}{3}(1+\star_4)k_{mn}(\check{v})
\end{eqnarray}
where
$\alpha_m(\check{v})\approx\frac{k_{mn}(\check{v})x^n}{r^4}$
near $r\rightarrow\infty$. Adding the
above two, what we get is exactly same as the ADM angular momentum
(\ref{ADM angular final}).


\subsection{$SU(2)$ 2-instantons: closed timelike curves and
charge bounds}

In this subsection we investigate the the case with $SU(2)$ gauge
group in detail. Since $d_{abc}=0$ for $SU(2)$, there is no
non-Abelian Chern-Simons term here. Since the single instanton is
basically given by the 't Hooft solution, which is quite special
rather than being generic, we concentrate on the case in which
instanton number $k$ is $2$. (For simplicity, we set $g=1$.)

The Yang-Mills 2-instanton for $SU(2)$ gauge group is completely given by
the so-called Jackiw-Nohl-Rebbi (JNR) solution \cite{jnr}. For $k=2$,
it is parameterized by \textit{three} positions $a_i$ ($i=0,1,2$) in
$\mathbb{R}^4$, and associated scales $\lambda_i$. The solution is given
as
\begin{equation}
  A^a_m=-\bar\eta^a_{\ mn}\partial_n \log H(x)\ ,\ \
  H\equiv\sum_{i=0}^2\frac{\lambda_i^{\ 2}}{|x-a_i|^2}\ ,
\end{equation}
where the anti-self-dual 't Hooft tensor $\bar\eta^a_{\ mn}$ is
defined as $\bar\sigma_{mn}\equiv i\bar\eta^a_{\ mn}\sigma^a$ (or
$\bar\eta^a_{\ bc}=\epsilon_{abc}$ and $\bar\eta^a_{\
b4}=-\delta^a_b$). One of the three scales $\lambda_i$ is
unphysical, since overall scaling of $H(x)$ does not affect the
gauge field $A^a_m$. Furthermore, as shown in \cite{jnr,at-ma}, one
of the twelve real parameters in $a_i$ is unphysical. To be more
precise, there is a unique circle in $\mathbb{R}^4$ passing through
the three points $a_i$. It is shown that moving the three points
along this circle with relative `speed' $\lambda_i^2$ can be undone
by a local gauge transformation. Thus one is left with $15-1-1=13$
independent parameters. Together with the 3 degrees in global gauge
orientation, they provide the complete parameterization of the
moduli space of $SU(2)$ 2-instantons.\footnote{From the general ADHM
solution, the above JNR solution can be obtained by appropriate
singular gauge transformation. See, for instance, \cite{ki-le} for
details.}

For convenience, we assume the scalars in vector multiplet live on
the coset, which is a symmetric space, explained in section 2. The
neutral and charged \cite{ki-le} scalars are given as
($C_\alpha\!=\!-1$ with $\alpha=1$ only, for this symmetric space
example)
\begin{equation}
  f^{-1}X_\alpha=h_\alpha+\frac{C_\alpha}{6}\partial^2\left(
  \log\left(\frac{s_0}{|x_0|^2|x_1|^2|x_2|^2}\right)-\log H\right)
  =h_\alpha+\frac{C_\alpha}{6}\left(\frac{\partial_m H\partial_m H}{H^2}
  -\sum_i\frac{4}{|x_i|^2}\right)
\end{equation}
and
\begin{equation}\label{k=2 charged scalar}
  \phi_a\frac{\sigma^a}{2}=\frac{1}{s_\Sigma H(x)}\left(
  \bar{Z}vZ+\frac{\mathcal{C}\sigma^a}{2}\bar\eta^a_{\ mn}
  \left(\frac{(x_0)^m}{|x_0|^2}\frac{(x_1)^n}{|x_1|^2}+
  \frac{(x_1)^m}{|x_1|^2}\frac{(x_2)^n}{|x_2|^2}+
  \frac{(x_2)^m}{|x_2|^2}\frac{(x_0)^n}{|x_0|^2}\right)\right)
\end{equation}
where $Z=\sigma_m\frac{s_i(x_i)^m}{|x_i|^2}\equiv \sigma_mZ_m$,
$v=v_a\frac{\sigma^a}{2}$, $s_i\equiv(\lambda_i)^2$,
$s_\Sigma=s_0+s_1+s_2$, $x_i=x-a_i$ and
\begin{equation}
  \mathcal{C}\equiv\frac{4v_a\eta^a_{\ mn}\left(\frac{}{}
  (a_0)_m(a_1)_n+(a_1)_m(a_2)_n+(a_2)_m(a_0)_n\right)}
  {(s_0s_1)^{-1}|a_0-a_1|^2+(s_1s_2)^{-1}|a_1-a_2|^2+
  (s_2s_0)^{-1}|a_2-a_0|^2}\ .
\end{equation}
From this expression one can obtain the function $f$. Assuming the above
symmetric space with $\mathcal{V}(x)=\frac{1}{2}X^1((X^2)^2-X^aX^a)$, one
finds
\begin{equation}
  f^{-3}=\frac{27}{2}(f^{-1}X_1)\left(\frac{}{}h_2^{\ 2}-\phi^a\phi^a\right)
  \ \ \ (>0\ \ {\rm everywhere})\ \ \
  \left(\frac{27}{2}h_1\left(\frac{}{}h_2^{\ 2}-v^av^a\right)=1\right)\ .
\end{equation}
Otherwise, we just understand that $f$ is given by the algebraic equation
$\frac{1}{6}C_{IJK}X^IX^JX^K=1$.

Now we turn to the 1-form $\omega_m$. Firstly, one can write
\begin{equation}\label{k=2 omega 1}
  -i{\rm tr}\left(\mathcal{J}_0(\mathcal{P}\partial_m\mathcal{P}
  -\partial_m\mathcal{P}\mathcal{P})\right)=-2{\rm tr}
  \left(\phi A_m\right)+i{\rm tr}\left(\bar{U}\mathcal{J}_0
  \partial_m U-\partial_m\bar{U}\mathcal{J}_0U\right)\ .
\end{equation}
After some computation, the second term can be written as
\begin{eqnarray}
  i{\rm tr}\left(\bar{U}\mathcal{J}
  \partial_m U-\partial_m\bar{U}\mathcal{J}U\right)&=&
  \frac{2v_a\eta^a_{\ np}}{s_\Sigma H(x)}
  \left(\frac{s_i(x_i)^n}{|x_i|^2}\right)\partial_m
  \left(\frac{s_j(x_j)^p}{|x_j|^2}\right)\\
  &&+\frac{\mathcal{C}}{s_\Sigma H(x)}\sum_{i=1}^3\epsilon^{ijk}
  \left(\frac{(x_j)^n}{|x_j|^2}\right)\partial_m
  \left(\frac{(x_k)^n}{|x_k|^2}\right)\ .\nonumber
\end{eqnarray}
With the following gauge field
\begin{equation}
  A_m=-\bar\eta^a_{\ mn}\frac{\sigma^a}{2}\frac{\partial_n H}{H}=
  +\frac{i}{2}\ \bar\sigma_{mn}\frac{\partial_n H}{H}
\end{equation}
and charged scalar solution, the first term becomes
\begin{eqnarray}
  \hspace*{-1cm}-2{\rm tr}(\phi A_m)
  &=&-\frac{1}{s_\Sigma}\ \partial_n\left(\frac{1}{H}\right)
  \left(\frac{}{}2v_a(Z_m\eta^a_{\ np}-Z_n\eta^a_{\ mp})Z_p+
  v_aZ_pZ_p\eta^a_{\ mn}\right)\nonumber\\
  &&-\frac{\mathcal{C}}{s_\Sigma}\ \partial_n
  \left(\frac{1}{H}\right)\left(\sum_i\epsilon^{ijk}
  \frac{(x_j)^m}{|x_j|^2}\frac{(x_k)^n}{|x_k|^2}-
  \frac{1}{2}\epsilon_{mnpq}
  \sum_i\epsilon^{ijk}\frac{(x_j)^p}{|x_j|^2}
  \frac{(x_k)^q}{|x_k|^2}\right)\nonumber
\end{eqnarray}
where we used
\begin{eqnarray}
  \bar\eta^a_{\ mn}\bar\eta^a_{\ pq}&=&\delta_{mp}\delta_{nq}-
  \delta_{mq}\delta_{np}-\epsilon_{mnpq}\\
  {\rm tr}(\bar\sigma_{(p}v\sigma_{q)}\bar\sigma_m\sigma_n)&=&
  2iv_a(\delta_{n(p}\eta^a_{\ q)m}-\delta_{m(p}\eta^a_{\ q)n})+
  iv_a\delta_{pq}\eta^a_{\ mn}\ .
\end{eqnarray}
Adding the two contributions, (\ref{k=2 omega 1}) becomes
\begin{eqnarray}
  &&\hspace{-1cm}-\frac{4v_a}{s_\Sigma}\eta^a_{\ mn}Z_n
  -\frac{2\mathcal{C}}{s_\Sigma H}\left(\frac{
  (a_0-a_1)^m}{|x_0|^2|x_1|^2}+\frac{(a_1-a_2)^m}{|x_1|^2|x_2|^2}
  +\frac{(a_2-a_0)^m}{|x_2|^2|x_0|^2}\right)
  -\bar\eta^a_{mn}\partial_n\phi^a(x)
\end{eqnarray}
where we used
\begin{equation}
  Z_m=\partial_m\left(\sum_i s_i\log|x_i|\right)\ \rightarrow\ \
  \partial_mZ_n-\partial_nZ_m=0\ .
\end{equation}
The second term appearing in (\ref{omega result}) is
\begin{equation}
  -i{\rm tr}\left(\bar{b}_\beta\mathcal{J}a_{\dot\alpha}
  \bar\sigma^{\dot\alpha\beta}_m\right)+c.c.=
  +\frac{\mathcal{C}}{s_\Sigma}\ \frac{|x_0|^2(a_1-a_2)^m+
  |x_1|^2(a_2-a_0)^m+|x_2|^2(a_0-a_1)^m}
  {s_0|x_1|^2|x_2|^2+s_1|x_2|^2|x_0|^2+s_2|x_0|^2|x_1|^2}\ ,
\end{equation}
so that $\omega_m$ itself simply becomes
\begin{equation}\label{k=2 omega final}
  \omega_m=-\frac{3}{2}\bar\eta^a_{mn}\partial_n\phi^a(x)-
  6\frac{v_a}{s_\Sigma}\eta^a_{\ mn}Z_n
  \ ,
\end{equation}
where the scalar is given as (\ref{k=2 charged scalar}).
One can explicitly check that (\ref{k=2 omega final}) is regular
everywhere including $x=a_i$, even if each term is not. This is just
re-confirming the regularity of our general solution.

To be concrete, let us consider the case where the three points
$a_i$ form an equilateral triangle on, say $x^1$-$x^2$ plane, with
scales $\lambda_i$ being all equal:
\begin{equation}
  a_0=(R,0,0,0)\ ,\ \ a_1=(-\frac{R}{2},\frac{\sqrt{3}R}{2},0,0)\ ,
  \ \ a_2=(-\frac{R}{2},-\frac{\sqrt{3}R}{2},0,0)\ ,\ \
  s_0=s_1=s_2=1\ .
\end{equation}
Then one obtains
\begin{equation}
  \mathcal{C}=\frac{2 v_3}{\sqrt{3}}
\end{equation}
The function $\det F(x)$ in this case hss $U(1)^2$ symmetry,
rotations on two 2-planes:
\begin{equation}
  \det F^{-1}(x)=|x_0|^2|x_1|^2|x_2|^2H=
  3\left((r^2+\rho^2+R^2)^2-R^2r^2\frac{}{}\right)\ ,
\end{equation}
where $r^2\equiv(x^1)^2+(x^2)^2$, $\rho^2\equiv(x^3)^2+(x^4)^2$. If
we take the scalar expectation to be $v_1=v_2=0$, which we do, this
symmetry of the gauge field becomes the symmetry of the full
solution. To see this, we first find that the gauge-invariant
combination $\phi^a\phi^a$ has this symmetry:
\begin{equation}
  \phi^a\phi^a=v^2-\frac{4v^2R^2}{9}\
  \frac{3(r^2+\rho^2)+5}{(r^2+\rho^2+R^2)^2-R^2r^2}+
  \frac{4v^2R^4}{3}\ \frac{\rho^4+\rho^2(r^2+2R^2)+R^2r^2+R^4}
  {((r^2+\rho^2+R^2)^2-R^2r^2)^2}\ .
\end{equation}
One can also obtain the 1-form $\omega_m$: defining $z\equiv x^1+ix^2$ and
$z^\prime\equiv x^3+ix^4$, one obtains after some algebra the following,
\begin{eqnarray}\label{circular omega}
  \omega_1-i\omega_2&=&\frac{2ivR^2\bar{z}
  (2(r^2+\rho^2+R^2)^2+R^2r^2)}{((r^2+\rho^2+R^2)^2-R^2r^2)^2}\\
  \omega_3-i\omega_4&=&-\frac{2ivR^2\bar{z}^\prime
  ((r^2+\rho^2+R^2)^2+2R^2r^2)}{((r^2+\rho^2+R^2)^2-R^2r^2)^2}\ ,
\end{eqnarray}
which also has symmetry under $U(1)^2$ rotations. The full geometry
is smooth everywhere.

Now we investigate if there is any closed timelike curves (CTC) in
the above geometry. We would check that there are no timelike
directions on the constant $t$ hyperspace. Pick up any unit vector
$N^m(x)$ in $\mathbb{R}^4$, that is $N^t=0$ and $N^mN^m=1$. The norm
of this vector is
\begin{equation}
  g_{\mu\nu}N^\mu N^\nu=f^{-1}\left(1-f^3(\omega_m N^m)^2\frac{}{}\right)
  \geq f^{-1}\left(1-f^3|\omega_m|^2\frac{}{}\right).
\end{equation}
Showing that the last expression never becomes negative will be
sufficient for proving the absence of CTC. To be precise, there
exists an ambiguity $\omega\rightarrow\omega+d\lambda$ associated
with shifting $t$ by $\lambda(x)$. However, we work with
(\ref{circular omega}) which will turn out to be enough to show
$f^3|\omega_m|^2<1$ everywhere. In fact we find
\begin{eqnarray}
  f^3|\omega_m|^2&=&\frac{(2vR^2)^2
  [(4r^2\!+\!\rho^2)(r^2\!+\!\rho^2\!+\!R^2)^4\!+\!4R^2r^2(r^2\!+\!\rho^2)
  (r^2\!+\!\rho^2\!+\!R^2)^2\!+\!R^4r^4(r^2\!+\!4\rho^2)]}
  {[(r^2+\rho^2+R^2)^2-R^2r^2]^4}\nonumber\\
  &&\times\left(h_1+\frac{1}{6}\partial^2\log
  \left[(r^2+\rho^2+R^2)^2-R^2r^2\right]\right)^{-1}\times\\
  &&\left(\frac{1}{h_1}+6v^2R^2\left(
  \frac{3(r^2+\rho^2)+5R^2}{(r^2+\rho^2+R^2)^2-R^2r^2}-
  \frac{3R^2(r^2\!+\!\rho^2\!+\!R^2)(\rho^2\!+\!R^2)}
  {[(r^2+\rho^2+R^2)^2-R^2r^2]^2}\right)\right)^{-1}<\frac{1}{4}
  \ .\nonumber
\end{eqnarray}
In particular, the upper bound $1$ is never attained. This confirms
that there are no CTC's in this $U(1)^2$-symmetric solutions. The
upper bound $\frac{1}{4}$ is asymptotically attained when
$v^2\rightarrow\infty$ and $h_1\rightarrow 0^+$, at $r=R$ and
$\rho=0$.

This absence of CTC in the above example may not be very surprising
since CTC appears as one tries to obtain an over-rotating solution.
For instance, one obtains the over-rotating BMPV black hole as one
takes to coefficient of the homogeneous solution for (\ref{final 3})
to be too large.  Since we only keep in $\omega_m$ the terms which
are not associated with singular sources, there seems to be no
degree in our solution to cause such an over-rotation.

Even if we believe that the absence of CTC can be true for our
general regular solutions, this seems to be hard for us to show
without symmetry, like the $U(1)^2$ isometry in the above example.
However, we shall provide an indirect evidence for this conjecture
for more general configurations. We show in the general 2-instanton
sector that the angular momentum has an upper bound given by other
charges. Especially, given the instanton number $k=2$ and electric
charge $q$, one finds that certain components of angular momenta are
maximized for the above $U(1)^2$ symmetric configurations.

For general 2-instantons, one obtains the following self-dual angular
momentum
\begin{equation}
  j_{mn}=8\pi^2(1+\star_4){\rm tr}\left(\varphi[a_m,a_n]\frac{}{}\right)
  =\frac{8\pi^2i}{s_\Sigma^{\ 2}}(1+\star_4)
  \left(\sum_i a_i\wedge a_{i+1}\right)_{mn}
  \frac{v^b\eta^b_{\ pq}\sum_i\left(a_i\wedge a_{i+1}\right)_{pq}}
  {\sum_i(s_is_{i+1})^{-1}|a_i-a_{i+1}|^2}\ .
\end{equation}
Note that, for 2-instantons, one can locate the three positions
$a_i$ on the 12 plane without losing generality. Defining
$j_{mn}\equiv\eta^a_{mn}j^a$, one finds that only $j^3$ is nonzero
and
\begin{equation}
  j^3=\frac{1}{4}\eta^3_{\ mn}j_{mn}=\frac{16\pi^2}{s_\Sigma^{\ 2}}
  \frac{{\rm vol}(\Delta(a_0a_1a_2))
  \left(\frac{}{}v^a\eta^a_{\ mn}\sum_i\left(a_i\wedge a_{i+1}\right)_{mn}
  \right)}{\sum_i(s_is_{i+1})^{-1}|a_i-a_{i+1}|^2}=
  \frac{4\pi^2v^3}{s_\Sigma^{\ 2}}\frac{(4{\rm vol}(\Delta(a_0a_1a_2)))^2}
  {\sum_i(s_is_{i+1})^{-1}|a_i-a_{i+1}|^2}
\end{equation}
where ${\rm vol}(\Delta(a_0a_1a_2))$ is the area of the triangle
made by three vectors $a_0$, $a_1$ and $a_2$. The electric charge is
given as \cite{ki-le}
\begin{eqnarray}
  v^aq^a&=&\frac{4\pi^2}{s_\Sigma^{\ 2}}
  \left(v^2\sum_i s_is_{i+1}|a_i-a_{i+1}|^2-
  \frac{\left(\frac{}{}v^a\eta^a_{\ mn}\sum_i\left(a_i\wedge a_{i+1}
  \right)_{mn}
  \right)^2}{\sum_i(s_is_{i+1})^{-1}|a_i-a_{i+1}|^2}\right)\nonumber\\
  &\geq&\frac{4\pi^2}{s_\Sigma^{\ 2}}\
  \frac{v^2\left(\frac{}{}\sum_i |a_i-a_{i+1}|^2\right)^2
  -\left(\frac{}{}v^a\eta^a_{\ mn}\sum_i\left(a_i\wedge a_{i+1}\right)_{mn}
  \right)^2}{\sum_i(s_is_{i+1})^{-1}|a_i-a_{i+1}|^2}
\end{eqnarray}
where we used the Schwarz inequality on the last line, which is
saturated if $s_0=s_1=s_2$. The above $v^aq^a$ and $|v|j^3$ (where
$|v|^2\equiv v^av^a$) satisfies the following inequality:
\begin{eqnarray}\label{sd inequality}
  \frac{s_\Sigma^{\ 2}}{4\pi^2}\left(v^aq^a-2|v|j^3\right)
  &\geq&\frac{v^2\left(\frac{}{}\sum_i |a_i-a_{i+1}|^2\right)^2
  -3v^2\left(\frac{}{}4{\rm vol}(\Delta(a_0a_1a_2))\frac{}{}\right)^2}
  {\sum_i(s_is_{i+1})^{-1}|a_i-a_{i+1}|^2}\nonumber\\
  &=&\frac{v^2\left(\frac{}{}\sum_i |a_i-a_{i+1}|^2\right)^2
  -3v^2\left(\frac{}{}2\sum_i |a_i-a_{i+1}|^2|a_i-a_{i-1}|^2
  -\sum_i|a_i-a_{i+1}|^4\right)}
  {\sum_i(s_is_{i+1})^{-1}|a_i-a_{i+1}|^2}\nonumber\\
  &=&2v^2\frac{\left(\frac{}{}(|a_{01}|^2\!-\!|a_{12}|^2)^2+
  (|a_{12}|^2\!-\!|a_{20}|^2)^2+(|a_{20}|^2\!-\!|a_{01}|^2)^2\right)}
  {\sum_i(s_is_{i+1})^{-1}|a_i-a_{i+1}|^2}\geq 0\ .
\end{eqnarray}
The two inequalities are saturated in the following cases,
respectively: (1) the first one if $s_0=s_1=s_2$ and $v^1=v^2=0$,
and (2) the second one if $|a_{01}|^2=|a_{12}|^2=|a_{20}|^2$.
Therefore we find $|j|\leq\frac{1}{2}\frac{v^aq^a}{|v|}$, which is
saturated by $U(1)^2$ invariant rings.

The anti-self-dual part of the angular momentum is given as
\begin{equation}
  \tilde{j}_{mn}=4\pi^2 i\
  {\rm tr}(\bar\omega^{\dot\alpha}v\omega_{\dot\beta})
  (\bar\sigma_{mn})^{\dot\beta}_{\ \dot\alpha}=
  \frac{4\pi^2 i}{s_\Sigma^{\ 2}}\left(s_0s_1\ \bar{a}_{01}va_{01}+
  s_1s_2\ \bar{a}_{12}va_{12}+s_2s_0\ \bar{a}_{20}va_{20}\frac{}{}
  \right)^{\dot\alpha}_{\ \dot\beta}
  (\bar\sigma_{mn})^{\dot\beta}_{\ \dot\alpha}\ .
\end{equation}
Here $a_{ij}\equiv (a_i-a_j)_m\sigma^m$. We define
$\tilde{j}_{mn}\equiv\bar\eta^a_{\ mn}\tilde{j}^a$, and again align
the vectors $a_{ij}$ on the 12 plane,
$a_{ij}=a_{ij}^1\sigma^1+a_{ij}^2\sigma^2$. Decomposing
$v=v_\parallel+v_\perp=v^3\frac{\sigma^3}{2}+(v^1\frac{\sigma^1}{2}
\!+\!v^2\frac{\sigma^2}{2})$, one finds
\begin{eqnarray}
  \tilde{j}_3&=&\frac{4\pi^2v_3}{s_\Sigma^{\ 2}}\left(\sum_i s_is_{i\!+\!1}
  |a_{i}\!-\!a_{i\!+\!1}|^2\right)\\
  \tilde{j}_1\frac{\sigma^1}{2}+\tilde{j}_2\frac{\sigma^2}{2}&=&
  -\frac{4\pi^2}{s_\Sigma^{\ 2}}\left(s_0s_1\ \bar{a}_{01}v_\perp
  a_{01}+s_1s_2\ \bar{a}_{12}v_\perp a_{12}+s_2s_0\ \bar{a}_{20}v_\perp
  a_{20}\frac{}{}\right)\nonumber\\
  &\equiv&-\frac{4\pi^2}{s_\Sigma^{\ 2}}\left(s_0s_1|a_{01}|^2
  v_\perp^{01}+s_1s_2|a_{12}|^2 v_\perp^{12}+s_2s_0|a_{20}|^2
  v_\perp^{20}\right)\label{asd perp}
\end{eqnarray}
where $v_\perp^{ij}\equiv\frac{\bar{a}_{ij}}{|a_{ij}|}
v_\perp\frac{a_{ij}}{|a_{ij}|}$. Regarding $v_\perp^{ij}$ as a
2-dimensional vector spanned by $\sigma^1$ and $\sigma^2$, it is a
rotation of $v_\perp$. From the structure of (\ref{asd perp}), one
finds that $(\tilde{j}_1)^2+(\tilde{j}_2)^2$ is maximized when all
$v_\perp^{ij}$ are parallel, which is possible when all $a_i$ lie on
the same line in $\mathbb{R}^4$. One finds
\begin{equation}
  \sqrt{(\tilde{j}_1)^2+(\tilde{j}_2)^2}\leq
  \frac{4\pi^2|v_\perp|}{s_\Sigma^{\ 2}}
  \left(\sum_i s_is_{i\!+\!1}|a_{i}\!-\!a_{i\!+\!1}|^2\right)\ .
\end{equation}
Therefore one obtains
\begin{equation}
  |\tilde{j}|=\sqrt{\sum_a\tilde{j}_a^{\ 2}}
  \leq \frac{4\pi^2|v|}{s_\Sigma^{\ 2}}
  \left(\sum_i s_is_{i\!+\!1}|a_{i}\!-\!a_{i\!+\!1}|^2\right)\ .
\end{equation}
This inequality is saturated if (i) $v_3=0$ and $a_{ij}$ all
parallel, or (ii) $v_\perp=0$. With this result, one finds that the
anti-self-dual angular momentum $\tilde{j}$ is also has an upper
bound given by the electric charge:
\begin{equation}\label{asd inequality}
  \frac{s_\Sigma^{\ 2}}{4\pi^2}
  \left(v^aq^a-\frac{2}{3}|v||\tilde{j}|\right)\geq
  \frac{|v|^2}{3}\sum_i s_is_{i+1}|a_i-a_{i+1}|^2-
  \frac{\left(\frac{}{}v^a\eta^a_{\ mn}\sum_i\left(a_i\wedge a_{i+1}
  \right)_{mn}
  \right)^2}{\sum_i(s_is_{i+1})^{-1}|a_i-a_{i+1}|^2}\geq 0\ ,
\end{equation}
where we applied the same inequalities used in (\ref{sd
inequality}). We therefore find
$|\tilde{j}|\leq\frac{3}{2}\frac{v^aq^a}{|v|}$. For all inequalities
used in the intermediate steps to be saturated, the configuration
should again satisfy $s_0=s_1=s_2$, $v_1=v_2=0$ and
$|a_{01}|=|a_{12}|=|a_{20}|$. Especially, both $|j|$ and
$|\tilde{j}|$ are bound by $\frac{v^aq^a}{|v|}$.\footnote{This
question was raised in \cite{ceh}, where a similar conclusion in a
slightly different setting was obtained.}

We suspect there could exist similar upper bound for general $SU(2)$
instantons with topological charge $k\geq 3$: perhaps similar to
what we found here, like $\frac{v^aq^a}{|v|}\geq c_{k}|j|$ and
$\frac{v^aq^a}{|v|}\geq \tilde{c}_k|\tilde{j}|$. We do not attempted
to explore it here, partly because we have not solved
(\ref{algebraic}) for $\varphi$ with general $k$, and also because
we cannot solve the ADHM constraint completely. For $k=1$, it is
known \cite{etz} that $j_{mn}=0$ while $|\tilde{j}|$ is proportional
to $\frac{q^av^a}{|v|}$. For $k=2$, one finds $j\neq 0$ in general,
but the upper bound for anti-self-dual part $|\tilde{j}|$ is still
larger than that for the self-dual part. The large $k$ expectation
is that the two bounds would be the same, namely
$\frac{\tilde{c}_k}{c_k}\rightarrow 1$ for $k\rightarrow\infty$
\cite{mnt,bho,ceh}. To see how such bounds behave for $k\geq 3$, if
they exist at all, one could restrict one's interest to the multi
JNR instanton of \cite{jnr}, where the ADHM data is also known
\cite{ki-le}. The matrix $\varphi$ is also obtained recently for
some values of $k\geq 3$ \cite{ckll}.

\section{Concluding remarks}

In this paper we studied supersymmetric solutions of 5 dimensional
$\mathcal{N}\!=\!1$ Yang-Mills-Einstein supergravity. We
systematically obtained explicit solutions to the differential
equations imposed on supersymmetric configurations based on ADHM
construction, modulo a set of algebraic conditions on the parameters
of the solutions. The solution carries topological charge, electric
charge and angular momentum. This gravitating dyonic instanton
solution is regular on the generic point of the instanton moduli
space.

We also checked the absence of CTC in the $U(1)^2$-invariant
solution carrying instanton charge $2$,  and conjectured the absence
for our general solution. It is indirectly supported in the general
2-instanton sector by showing the existence of un upper bound for
angular momenta in $\mathbb{R}^4$. It will be interesting to further
explore it.

In the truncated $\mathcal{N}=2$ model, the dyonic instantons in 5
dimensional super-Yang-Mills theory have been argued to be
supertubes, configurations carrying suitable dipole charges and
expanding into `tubular' or `ring-like' shapes in space. We find
further evidence for this interpretation in the theory with $SU(2)$
gauge group, by showing that both self-dual and anti-self-dual
components of the angular momentum are maximized for circular
configurations with $U(1)^2$ symmetry in the 2-instanton sector.

In the theory with non-Abelian Chern-Simons term, even the gauge
theory soliton needs further study. There we find that our
configuration has non-zero electric charge even if the adjoint
scalars take zero VEV, leaving $SU(N)$ gauge symmetry unbroken. The
equation (\ref{algebraic}) for $\varphi$ appearing in the scalar
solution has a natural interpretation as a non-dynamical auxiliary
degree in the matrix quantum mechanics describing the dynamics of
$k$-instanton moduli $\omega_{\dot\alpha}$ and $a_n$: the latter
model arises either as the moduli space approximation or as
describing open strings degrees attached to $D0$-$D4$ branes. When
$c=0$, from the latter viewpoint, since there is a $U(k)$ gauge
symmetry on $k$ stacks of $D0$ branes, one introduces a gauge field
$A_0$ and its superpartner scalar, which we call $\varphi$, living
on the worldline. The equation of motion for $\varphi$ is exactly
(\ref{algebraic}) with $c=0$. We managed to find a deformation of
this matrix model with the parameter $c\neq 0$, preserving $8$
supersymmetries, which yields (\ref{algebraic}) as the equation of
motion for $\varphi$, and further reproduces
(\ref{cs-single-charge}) in the single instanton
sector.\footnote{This is in progess in collaboration with Ki-Myeong
Lee.} It should be interesting to understand this finding more
physically.

In a broader perspective, one could extend the study of non-Abelian
supersymmetric solutions to other gauged supergravity theories. For
example, if one gauges both $U(1)\subset SU(2)_R$ as well as an
isometry on scalar manifolds, the resulting theory has nonzero
scalar potential. Gauged supergravity with $\mathcal{N}=2$ (16 real)
or $\mathcal{N}=4$ (32 real) supersymmetry is another direction. In
a theory where a subgroup of $SU(2)_R$ is gauged, the global
$SU(2)_R$ symmetry is broken by picking up a $U(1)$ subgroup.
Related to this, the hyper-Kahler structure on the base space that
we got should be relaxed \cite{ga-gu1}, which could render the
system richer and/or more complicated.

\vskip 0.7cm

\hspace*{-0.6cm}{\bf\large Acknowledgements}

\hspace*{-0.6cm}We are grateful to Ki-Myeong Lee and David Tong for
helpful discussions. We would also like to thank Kevin Goldstein,
Choonkyu Lee, Sangmin Lee and Daniel Waldram for conversations. S.L.
is supported in part by the Korea Research Foundation Grant
R14-2003-012-01001-0.


\appendix

\section{Derivation of the ADHM solutions}

In this appendix, we solve the differential conditions (\ref{final
2}) and (\ref{final 3}) using ADHM technique. For convenience, we
set the gauging parameter $g=1$ here, which can be recovered easily.

\subsection{Adjoint scalar solution}

In this subsection, we derive the solution of the covariant Laplace
equation with a source term coming from non-Abelian Chern-Simons
coupling:
\begin{equation}
  \mathcal{D}^2(f^{-1}X_a)=\frac{c}{12}d_{abc}\Theta^b_{mn}\Theta^c_{mn}\ .
\end{equation}
Alternatively, in $N\times N$ matrix notation, one may first solve
an auxiliary equation
\begin{equation}
  \mathcal{D}^2\Phi=\frac{c}{24}\Theta_{mn}\Theta_{mn}\ \ \
  (\{T^a,T^b\}=\frac{1}{N}\delta^{ab}{\bf 1}_{N}+4d^{abc}T^c)\ .
\end{equation}
Since there is an overall $U(1)$ part, whose solution is given by
the Osborn's formula
\begin{equation}
  {\rm tr}\Phi=\frac{c}{24}\partial^2\log(\det F(x))\ ,
\end{equation}
$f^{-1}X_a$ is obtained from $\Phi$ as
\begin{equation}
  f^{-1}X_aT^a=\Phi-\frac{1}{N}({\rm tr}\Phi){\bf 1}_N
  =\Phi-\frac{c}{24N}\partial^2\log(\det F(x))\ {\bf 1}_N\ .
\end{equation}

Using $(\sigma_{mn})_\alpha^{\ \beta}(\sigma_{mn})_\gamma^{\
\delta}\!=\!-4\left(\delta_\alpha^{\ \delta}\delta^\beta_{\ \gamma}+
\epsilon_{\alpha\gamma}\epsilon^{\beta\delta}\right)$, one obtains
from $\Theta_{mn}=2i\bar{U}b(\sigma_{mn}F)\bar{b}U$ the following:
\begin{equation}
  \Theta_{mn}\Theta_{mn}=
  +16\left(\bar{U}b^\alpha F \bar{b}_\beta\mathcal{P}b^\beta
  F\bar{b}_\alpha U+\bar{U}b^\alpha F \bar{b}_\beta\mathcal{P}
  b_\alpha F\bar{b}^\beta U\right)\ .
\end{equation}
As a first trial, we compute $\mathcal{D}^2(\bar{U}\mathcal{J}_1U)$
with $\mathcal{J}_1\equiv b^\alpha F\bar{b}_\alpha$. The general
expression in \cite{dhkm} is
\begin{eqnarray}
  \mathcal{D}^2(\bar{U}\mathcal{J}_1U)&=&-4\bar{U}\{b^\alpha F\bar{b}_\alpha,
  \mathcal{J}_1\}U+4\bar{U}b^\alpha
  F\bar\Delta^{\dot\alpha}\mathcal{J}_1\Delta_{\dot\alpha}F\bar{b}_\alpha
  U\nonumber\\
  &&+\bar{U}\partial^2\mathcal{J}_1U-2\bar{U}b^\alpha F
  \sigma_{n\alpha\dot\alpha}\bar\Delta^{\dot\alpha}\partial_n\mathcal{J}_1
  U-2\bar{U}\partial_n\mathcal{J}_1\Delta_{\dot\alpha}
  \bar\sigma_n^{\dot\alpha\alpha}F\bar{b}_\alpha U\ .
\end{eqnarray}
Inserting $\mathcal{J}_1=b^\alpha F\bar{b}_\alpha$, one obtains
\begin{eqnarray}\label{non-ab cs trial}
  \mathcal{D}^2(\bar{U}\mathcal{J}_1U)&=&-8\bar{U}
  (b^\alpha F\bar{b}_\alpha)^2U+4\bar{U}b^\alpha
  F\bar\Delta^{\dot\alpha}b^\beta F\bar{b}_\beta\Delta_{\dot\alpha}
  F\bar{b}_\alpha U\nonumber\\
  &&-4\bar{U}b^\alpha F\bar{b}_\beta \mathcal{P}b^\beta F\bar{b}_\alpha U
  +8\bar{U}b^\alpha F\bar\Delta^{\dot\alpha}b^\beta F\bar{b}_\alpha
  \Delta_{\dot\alpha}F\bar{b}_\beta U
\end{eqnarray}
where $\mathcal{P}=U\bar{U}$. Here we used
\begin{eqnarray}
  \partial^2\mathcal{J}_1&=&-4b^\alpha F \bar{b}_\beta\mathcal{P}b^\beta
  F\bar{b}_\alpha\\
  \partial_n\mathcal{J}_1&=&\left\{
  \begin{array}{c}-b^\alpha F\bar\sigma_n^{\dot\beta\beta}
  \bar{b}_\beta\Delta_{\dot\beta} F \bar{b}_\alpha\\
  -b^\alpha F\bar\Delta^{\dot\beta}b^\beta\sigma_{n\beta\dot\beta}
  F\bar{b}_\alpha\end{array}\right.
\end{eqnarray}
and
$\sigma_{n\alpha\dot\alpha}\bar\sigma_n^{\dot\beta\beta}=2\delta_\alpha^{\
\beta}\delta_{\dot\alpha}^{\ \dot\beta}$. We try to massage the
second and fourth terms:
\begin{eqnarray}
  4\bar{U}b^\alpha
  F\bar\Delta^{\dot\alpha}b^\beta F\bar{b}_\beta\Delta_{\dot\alpha}
  F\bar{b}_\alpha U&=&4\bar{U}b^\alpha F\bar{b}_\beta(1-\mathcal{P})
  b^\beta F\bar{b}_\alpha U=
  8\bar{U}(b^\alpha F\bar{b}_\alpha)^2U
  -4\bar{U}b^\alpha F\bar{b}_\beta\mathcal{P}
  b^\beta F\bar{b}_\alpha U\nonumber\\
  8\bar{U}b^\alpha F\bar\Delta^{\dot\alpha}b^\beta F\bar{b}_\alpha
  \Delta_{\dot\alpha}F\bar{b}_\beta U&=&8\bar{U}b^\alpha
  F\bar{b}_\beta(1-\mathcal{P})b_\alpha F\bar{b}^\beta U=
  8\bar{U}(b^\alpha F\bar{b}_\alpha)^2U-8\bar{U}b^\alpha
  F\bar{b}_\beta\mathcal{P}b_\alpha F\bar{b}^\beta U\nonumber
\end{eqnarray}
where we used
$\bar\Delta_{\dot\alpha}b_\alpha=\bar{b}_\alpha\Delta_{\dot\alpha}$.
Inserting these back to (\ref{non-ab cs trial}), one obtains
\begin{eqnarray}
  \mathcal{D}^2(\bar{U}\mathcal{J}_1U)&=&
  8\bar{U}(b^\alpha F\bar{b}_\alpha)^2U
  -8\bar{U}b^\alpha F\bar{b}_\beta \mathcal{P}b^\beta F\bar{b}_\alpha U
  -8\bar{U}b^\alpha
  F\bar{b}_\beta\mathcal{P}b_\alpha F\bar{b}^\beta U\nonumber\\
  &=&8\bar{U}(b^\alpha F\bar{b}_\alpha)^2U-\frac{1}{2}
  \Theta_{mn}\Theta_{mn}\ .
\end{eqnarray}
Therefore, $\Phi$ has to satisfy
\begin{equation}\label{na-CS-trial}
  \mathcal{D}^2\left(\Phi+\frac{c}{12}\bar{U}b^\alpha F\bar{b}_\alpha U
  \right)=+\frac{2c}{3}\bar{U}b^\alpha F^2\bar{b}_\alpha U\ .
\end{equation}
The last equation can be solved by generalizing the ansatz taken in
\cite{dhkm} to solve the covariant Laplace equation. We try
\begin{equation}
  \Phi+\frac{c}{12}\bar{U}b^\alpha F\bar{b}_\alpha
  U=\bar{U}\left(\begin{array}{cc}v&0\\0&\varphi\otimes{\bf 1}_2
  \end{array}\right)U\equiv\bar{U}\mathcal{J}_0U\ ,
\end{equation}
where $v$ is the asymptotic value of $X_aT^a$, and $\varphi$ is a
constant matrix to be determined. Plugging this ansatz in
(\ref{na-CS-trial}) and following the computation (C.31) of
\cite{dhkm}, one obtains
\begin{equation}
 4\bar{U}b^\alpha F\left(
 -{\bf L}\varphi+\bar\omega^{\dot\alpha}v\omega_{\dot\alpha}\right)
 F\bar{b}_\alpha U=\frac{2c}{3}\bar{U}b^\alpha F^2\bar{b}_\alpha U
\end{equation}
where ${\bf
L}\varphi=\frac{1}{2}\{{\bar\omega}^{\dot\alpha}\omega_{\dot\alpha},\varphi\}+
[a_n,[a_n,\varphi]]$. This equation is solved if one demands
\begin{equation}\label{alg-modified}
  {\bf
  L}\varphi=\bar\omega^{\dot\alpha}v\omega_{\dot\alpha}-
  \frac{c}{6}{\bf 1}_k\ ,
\end{equation}
which is solvable since ${\bf L}$ is generically invertible. The
final answer is
\begin{equation}\label{scalar sol}
  f^{-1}X_aT^a=
  \bar{U}\left(\begin{array}{cc}v&0\\0&\varphi-\frac{c}{12}F(x)
  \end{array}\right)U-\frac{c}{24N}\partial^2\log(\det F(x))
  \ {\bf 1}_N
\end{equation}
with (\ref{alg-modified}).

\subsection{The 1-form $\omega_m$}

Here we derive the solution of (\ref{final 3}), where the scalar on
the right hands side is given by (\ref{scalar sol}).

Again as a first trial, we would like to compute the action of
$\frac{1+\ast_4}{2}d$ on the 1-form ${\rm
tr}\left(\mathcal{J}_0\mathcal{P}\partial_m\mathcal{P}\right)$.
Using the following identities \cite{dhkm},
\begin{eqnarray}
  \partial_m F&=&\left\{
  \begin{array}{l}-F\bar\sigma_m^{\dot\alpha\alpha}\bar{b}_\alpha
  \Delta_{\dot\alpha}F\\-F\bar\Delta^{\dot\alpha}b^\alpha
  \sigma_{n\alpha\dot\alpha}F\end{array}\right.\\
  \partial_m\mathcal{P}&=&-\Delta_{\dot\alpha}F
  \bar\sigma_m^{\dot\alpha\alpha}\bar{b}_\alpha\mathcal{P}-\mathcal{P}
  b^\alpha\sigma_{n\alpha\dot\alpha}F\bar\Delta^{\dot\alpha}\ ,
\end{eqnarray}
one obtains
\begin{equation}\label{omega inter 1}
  \frac{1+\ast_4}{2}\
  {\rm tr}\left(\mathcal{J}_0\partial_{[m}
  \mathcal{P}\partial_{n]}\mathcal{P}\right)=
  \frac{1+\ast_4}{2}{\rm tr}\left(\mathcal{J}
  \left(\mathcal{P}b^\alpha(\sigma_{mn})_\alpha^{\ \beta}
  F\bar{b}_\beta\mathcal{P}+
  \Delta_{\dot\alpha}F\bar\sigma_{[m}^{\ \dot\alpha\alpha}
  \bar{b}_\alpha\mathcal{P}b^\beta\sigma_{n]\beta\dot\beta}
  F\bar\Delta^{\dot\beta}\right)\frac{}{}\right)\ .
\end{equation}
To treat the second term, one needs
\begin{equation}\label{pauli fierz}
  (\sigma_{m})_{\alpha\dot\beta}(\bar\sigma_{n})^{\ \dot\gamma\delta}=
  \frac{1}{2}\delta_{mn}\delta_\alpha^{\ \delta}
  \delta_{\dot\beta}^{\ \dot\delta}-\frac{1}{2}\delta_\alpha^{\ \delta}
  (\bar\sigma_{mn})^{\dot\gamma}_{\ \dot\beta}+\frac{1}{2}
  (\sigma_{mn})_{\alpha}^{\ \delta}\delta^{\dot\gamma}_{\ \dot\beta}
  -\frac{1}{2}(\bar\sigma_{mp})^{\dot\gamma}_{\ \dot\beta}
  (\sigma_{pn})_\alpha^{\ \delta}\ .
\end{equation}
The last term on the right hand side of (\ref{pauli fierz}) is zero
after anti-symmetrizing $mn$ indices. We thus find a useful identity
\begin{equation}\label{gamma self-dual}
  \frac{1+\ast_4}{2}\
  (\sigma_{[m})_{\alpha\dot\beta}(\bar\sigma_{n]})^{\ \dot\gamma\delta}
  =\frac{1}{2}(\sigma_{mn})_{\alpha}^{\ \delta}
  \delta^{\dot\gamma}_{\ \dot\beta}\ .
\end{equation}
We also need the following property \cite{dhkm},
\begin{equation}
  \bar\Delta^{\dot\alpha}\mathcal{J}_0\Delta_{\dot\alpha}=
  \bar\omega^{\dot\alpha}v\omega_{\dot\alpha}-{\bf L}\varphi+
  \{\varphi,F^{-1}\}=\frac{c}{6}{\bf 1}_k+\{\varphi,F^{-1}\}\ .
\end{equation}
Using these, the quantity inside the $\frac{1+\star_4}{2}$ projector
of (\ref{omega inter 1}) can be written as
\begin{eqnarray}\label{omega inter 2}
  &&{\rm tr}\left(\mathcal{J}_0
  \mathcal{P}b^\alpha(\sigma_{mn})_\alpha^{\ \beta}
  F\bar{b}_\beta\mathcal{P}\right)+
  \frac{1}{2}{\rm tr}\left((\bar\Delta^{\dot\alpha}\mathcal{J}_0
  \Delta_{\dot\alpha})F\bar{b}_\alpha\mathcal{P}
  (\sigma_{nm})_\beta^{\ \alpha}b^\beta
  F\frac{}{}\right)\\
  &&={\rm tr}\left(\left(\bar{U}\mathcal{J}_0
  U\right)\left(\frac{}{}\bar{U}b^\alpha(\sigma_{mn})_\alpha^{\ \beta}
  F\bar{b}_\beta U\right)\right)+
  \frac{1}{2}{\rm tr}\left(\{\varphi,F\}\ \bar{b}_\alpha
  \mathcal{P}(\sigma_{nm})_\beta^{\ \alpha}b^\beta
  \frac{}{}\right)-\frac{c}{12}{\rm tr}\left(b^\alpha(\sigma_{mn})_\alpha
  ^{\ \beta}F^2\bar{b}_\beta\mathcal{P}\frac{}{}\right)\nonumber\\
  &&=-\frac{i}{2}{\rm tr}\left(\frac{}{}(f^{-1}X)F_{mn}\right)+
  \frac{1}{2}{\rm tr}\left(\{\mathcal{P},\mathcal{J}_0\}\
  b^\beta(\sigma_{nm})_\beta^{\ \alpha}
  F\bar{b}_\alpha\frac{}{}\right)-\frac{c}{12}
  {\rm tr}\left(b^\alpha(\sigma_{mn})_\alpha
  ^{\ \beta}F\bar{b}_\beta(1-\mathcal{P})\mathcal{J}_1\mathcal{P}
  \frac{}{}\right)\ ,\nonumber
\end{eqnarray}
where $f^{-1}X\equiv f^{-1}X_aT^a$ is given as (\ref{scalar sol}).
The first term is what we need on the right hand side of (\ref{final
3}). We shall explain how to deal with the other two terms below.

First we show that the second term on the last line of (\ref{omega
inter 2}) can be arranged to take the form $\frac{1+\ast_4}{2}\
d(\cdots)$. First, $\mathcal{P}$ appearing in this term can be
replaced by
$-(1-\mathcal{P})=-\Delta_{\dot\alpha}F\bar\Delta^{\dot\alpha}$,
since there is $\bar{b}_\alpha\mathcal{J}b^\beta=\varphi\
\delta_\alpha^\beta$ in the subtracted term, from which one finds
$(\sigma_{mn})_\alpha^{\ \alpha}=0$. Thus we consider
\begin{equation}
  -\frac{1}{2}{\rm tr}\left(\{
  \Delta_{\dot\alpha}F\bar\Delta^{\dot\alpha},\mathcal{J}\}\
  b^\beta(\sigma_{nm})_\beta^{\ \alpha}
  F\bar{b}_\alpha\frac{}{}\right)\ .
\end{equation}
We use (\ref{gamma self-dual}) to rewrite this term as
\begin{equation}
  -\frac{1+\ast_4}{2}\ {\rm tr}\left(\{
  \Delta_{\dot\alpha}F\bar\Delta^{\dot\beta},\mathcal{J}\}\
  b^\beta(\sigma_{[n})_{\beta\dot\beta}
  F(\bar\sigma_{m]})^{\ \dot\alpha\alpha}
  \bar{b}_\alpha\frac{}{}\right)=
  -\frac{1+\ast_4}{2}\ {\rm tr}\left(\{
  \Delta_{\dot\alpha}F\bar\Delta^{\dot\beta},\mathcal{J}\}\
  \partial_{[n}\Delta_{\dot\beta}
  F\partial_{m]}\bar\Delta^\alpha\frac{}{}\right)\ .
\end{equation}
Using the fact
$(\partial_m\bar\Delta^{\dot\alpha})\Delta_{\dot\alpha}=
\bar\Delta^{\dot\alpha}(\partial_m\Delta_{\dot\alpha})=
\partial_mF^{-1}$, this can be written as
\begin{equation}
  +\frac{1+\ast_4}{2}\ {\rm tr}\left(
  \mathcal{J}\Delta_{\dot\alpha}
  \partial_{[n}F\partial_{m]}\bar\Delta^{\dot\alpha}+
  \mathcal{J}\partial_{[n}\Delta_{\dot\alpha}
  \partial_{m]}F\bar\Delta^{\dot\alpha}
  \frac{}{}\right)\ .
\end{equation}
Each term inside the $\frac{1+\ast_4}{2}$ projector is exact. The
first term is
\begin{eqnarray}
  {\rm tr}\left(
  \mathcal{J}\Delta_{\dot\alpha}
  \partial_{[n}F\partial_{m]}\bar\Delta^{\dot\alpha}
  \frac{}{}\right)&=&{\rm tr}\left(\mathcal{J}
  (a_{\dot\alpha}\!+\!b^\alpha\sigma_{p\alpha\dot\alpha}x^p)
  \partial_{[n}F\bar\sigma_{m]}^{\ \dot\alpha\beta}
  \bar{b}_\beta\frac{}{}\right)\\
  &=&-\partial_{[m}{\rm tr}\left(\bar{b}_\beta\mathcal{J}
  a_{\dot\alpha}F\bar\sigma_{n]}^{\ \dot\alpha\beta}
  +2x_{n]}\varphi F\frac{}{}\right)\ ,\nonumber
\end{eqnarray}
and similarly the second term is
\begin{equation}
  {\rm tr}\left(
  \mathcal{J}\partial_{[n}\Delta_{\dot\alpha}
  \partial_{m]}F\bar\Delta^{\dot\alpha}
  \frac{}{}\right)=
  +\partial_{[m}{\rm tr}\left(\bar{a}^{\dot\alpha}\mathcal{J}
  b^\beta F\sigma_{n]\beta\dot\alpha}
  +2x_{n]}\varphi F\frac{}{}\right)\ .
\end{equation}
Collecting all, one obtains the following expression
\begin{equation}
  \frac{1}{2}{\rm tr}\left(\{\mathcal{P},\mathcal{J}\}\
  b^\beta(\sigma_{nm})_\beta^{\ \alpha}
  F\bar{b}_\alpha\frac{}{}\right)=\frac{1+\ast_4}{2}\partial_{[m}
  {\rm tr}\left(\left(\bar{a}^{\dot\alpha}\mathcal{J}
  b^\beta \sigma_{n]\beta\dot\alpha}-
  \bar{b}_\beta\mathcal{J}
  a_{\dot\alpha}\bar\sigma_{n]}^{\ \dot\alpha\beta}
  \right)F\right)\ .
\end{equation}
The expression inside the derivative can simply be rewritten as
\begin{equation}
  {\rm tr}
  \left(\left(\bar{b}_\beta\mathcal{J}
  a_{\dot\alpha}\bar\sigma_{n}^{\ \dot\alpha\beta}-
  \bar{a}^{\dot\alpha}\mathcal{J}
  b^\beta \sigma_{n\beta\dot\alpha}\right)F\frac{}{}\right)=
  2{\rm tr}_k\left([\varphi,a_m]F\right)\ ,
\end{equation}
where the lower $2k\times 2k$ block of $a_{\dot\alpha}$ is written
as $a_n\sigma^n_{\alpha\dot\alpha}$. This completes the analysis of
the second term of (\ref{omega inter 2}).

As the final step, we try to write the third term of (\ref{omega
inter 2}) in the form $\frac{1+\star_4}{2}d(\cdots)$. We first note
that this last term can be written in either of the following ways:
\begin{equation}\label{omega inter 3}
  \mathcal{O}_{mn}\equiv
  {\rm tr}\left(\frac{}{}b^\alpha(\sigma_{mn})_\alpha^{\ \beta}
  F\bar{b}_\beta(1-\mathcal{P})\mathcal{J}_1\mathcal{P}\right)=
  {\rm tr}\left(\frac{}{}b^\alpha(\sigma_{mn})_\alpha^{\ \beta}
  F\bar{b}_\beta\mathcal{P}\mathcal{J}_1(1-\mathcal{P})\right)\ .
\end{equation}
One can write this term in another way by using the following
identity,
\begin{equation}\label{pauli fierz 2}
  (\sigma_{mn})_\alpha^{\ \beta}\delta_\gamma^{\ \delta}=
  \frac{1}{2}(\sigma_{mn})_\alpha^{\ \delta}\delta_{\gamma}^{\
  \beta}+\frac{1}{2}(\sigma_{mn})_\gamma^{\ \beta}\delta_{\alpha}^{\
  \delta}-\frac{1}{4}\left(\frac{}{}(\sigma_{mp})_\alpha^{\ \delta}
  (\sigma_{np})_\gamma^{\ \beta}-(\sigma_{np})_\alpha^{\ \delta}
  (\sigma_{mp})_\gamma^{\ \beta}\right)\ .
\end{equation}
Applying this idendity to the latter form in (\ref{omega inter 3}),
one obtains
\begin{eqnarray}\label{co-closed}
  \mathcal{O}_{mn}&=&{\rm tr}\left(\frac{}{}b^\alpha
  (\sigma_{mn})_\alpha^{\ \beta}F^2\bar{b}_\beta(\Delta_{\dot\alpha}F
  \bar\Delta^{\dot\alpha})\right)-{\rm tr}\left(\frac{}{}b^\alpha
  (\sigma_{mn})_\alpha^{\ \beta}F\bar{b}_{\beta}
  (\Delta_{\dot\alpha}F\bar\Delta^{\dot\alpha})b^\gamma F\bar{b}_\gamma
  (\Delta_{\dot\beta}F\bar\Delta^{\dot\beta})\right)\nonumber\\
  &=&\frac{1+\star_4}{2}{\rm tr}\left(\frac{}{}2\partial_{[m}F^{-1}F^2
  \partial_{n]}F^{-1} F\right)+{\rm tr}\left(\frac{}{}
  \Delta_{\dot\alpha}F\bar\Delta^{\dot\beta}
  b^\alpha(\sigma_{mn})_\alpha^{\ \beta}F\bar{b}_\beta\Delta_{\dot\beta}
  F\bar\Delta^{\dot\alpha}b^\gamma F\bar{b}_\gamma\right)\nonumber\\
  &=&\frac{1+\star_4}{2}\ 2{\rm tr}\left(\frac{}{}
  \partial_{[m}F^{-1}F^2
  \partial_{n]} F^{-1} F+(F\partial_{[m} F^{-1}F\partial_{n]} F^{-1}F)
  \bar{b}_\alpha\Delta_{\dot\alpha}F\bar\Delta^{\dot\alpha}b^\alpha\right)\\
  &=&2\ \frac{1+\star_4}{2}\ {\rm tr}\left(\frac{}{}
  (F\partial_{[m}F^{-1}F\partial_{n]}F^{-1}F)(1-\bar{b}_\alpha\mathcal{P}
  b^\alpha)\right)\nonumber\\
  &=&+\frac{1+\star_4}{2}\ \frac{1}{2}{\rm tr}\left(
  \frac{}{}\partial_{[m}F^{-1}F\partial_{n]}F^{-1}(\partial^2 F+4 F^2)
  \right)\equiv\frac{1+\star_4}{2}\ \varrho_{mn}\ .\nonumber
\end{eqnarray}
We applied the above identity (\ref{pauli fierz 2}) to the second
term on the first line, and also used $\partial^2
F=-4F\bar{b}_\alpha\mathcal{P}b^\alpha F$ \cite{dhkm} one the 4th
line. Here we note that, inside the projector $\frac{1+\star_4}{2}$,
proving that the 2-form $\varrho_{mn}$ is co-exact is also fine for
our purpose. Namely we try to write $\varrho=d^\dag\omega^{(3)}$
with certain 2-form $\omega^{(3)}$, where $d^\dag\equiv\star_4
d\star_4$. The following 3-form
\begin{eqnarray}
  \Lambda_{mnp}&=&{\rm tr}
  \left(\partial_{[m}F^{-1}F\partial_{n]}F^{-1}\partial_pF\right)
  =-{\rm tr}\left(\partial_{[m}F^{-1}F\partial_{n]}F^{-1}
  F\partial_pF^{-1}F\right)
\end{eqnarray}
turns out to be helpful. Since the three indices $mnp$ are symmetric
under cyclic permutations, anti-symmetrizing $mn$ guarantees that
the indices are totally anti-symmetric. Acting $\partial_p$ on this
3-form, and using $\partial_m\partial_nF^{-1}=2\delta_{mn}{\bf
1}_{k}$, one obtains
\begin{equation}
  \partial_p\Lambda_{mnp}=2\varrho_{mn}\ .
\end{equation}
Inside the projector $\frac{1+\star_4}{2}$, one can write
\begin{equation}
  \frac{1+\star_4}{2}\partial_p\Lambda_{mnp}
  =\frac{1+\star_4}{2}\left(\star_4 d\lambda^{(1)}\right)_{mn}
  =\frac{1+\star_4}{2}\left(d\lambda^{(1)}\right)_{mn}
\end{equation}
where $\lambda^{(1)}\equiv\star_4\Lambda$.

Collecting all, the 1-form $\omega_m$ is given as
\begin{equation}\label{omega cs}
  \omega_m=-3i\ {\rm tr}\left(\mathcal{J}_0\frac{\mathcal{P}\partial_m
  \mathcal{P}-\partial_m\mathcal{P}\mathcal{P}}{2}+
  2[\varphi,a_m]F-\frac{c}{72}\epsilon_{mnpq}
  \partial_{n}F^{-1}F\partial_{p}F^{-1}
  F\partial_qF^{-1}F\right)
\end{equation}
where the traces are either over $N+2k$ or $k$ dimensional matrices,
and $\varphi$ appearing in $\mathcal{J}_0$ satisfies
(\ref{alg-modified}).

\section{Summary of the properties of spinor bilinear}

In this appendix we summarize the algebraic and differential
conditions satisfied by the differential forms constructed from the
Killing spinor bilinear. These are nearly the same as the conditions
in Maxwell-Einstein supergravity. We follow the notations of
\cite{gu-sa}.

The algebraic conditions following from the Fierz identity are
\begin{eqnarray}
  V_\mu V^\mu&=&f^2\label{alg 1}\\
  J^i\wedge J^j&=&-2\delta_{ij}f\star V\label{alg 2}\\
  i_V J^i&=&0\label{alg 3}\\
  i_V \star J^i&=&-f J^i\label{alg 4}\\
  J^i_{\rho\mu}{J^j}^\rho_{\ \nu}&=&\delta_{ij}\left(f^2\eta_{\mu\nu}-
  V_\mu V_\nu\right)+\epsilon_{ijk}fJ^k_{\mu\nu}\ .\label{alg 5}
\end{eqnarray}
The differential conditions that one obtains from the gravitino Killing
spinor equation are
\begin{eqnarray}
  df&=&-i_V\left(X_IF^I\right)\label{diff 1}\\
  \nabla_{(\mu}V_{\nu)}&=&0\label{diff 2}\\
  dV&=&2fX_IF^I+X_I\star\left(F^I\wedge V\right)\label{diff 3}\\
  \nabla_\mu J^i_{\nu\rho}&=&-\frac{1}{2}X_I\left(2{F^I}_\mu^{\ \sigma}
  \left(\star J^i\right)_{\sigma\nu\rho}-2{F^I}_{[\nu}^{\ \ \sigma}
  \left(\star J^i\right)_{\rho]\mu\sigma}+
  \eta_{\mu[\nu}{F^I}^{\sigma\tau}\left(\star J^i\right)_{\rho]\sigma\tau}
  \right)\label{diff 4}
\end{eqnarray}
These conditions are same as the results in \cite{gu-sa}, except
that we are setting $\chi=0$ (a parameter in their scalar potential)
in their formulae. The conditions coming from the gaugino Killing
spinor equation is slightly different to \cite{gu-sa}. Contracting
this equation with $\bar\epsilon^j$, one obtains
\begin{eqnarray}
  V^\mu D_\mu X_I&=&0\label{diff 5}\\
  \left(\frac{1}{4}Q_{IJ}-\frac{3}{8}X_IX_J\right)F^J_{\mu\nu}
  {J^i}^{\mu\nu}&=&0\ .\label{diff 6}
\end{eqnarray}
Contracting it with $\bar\epsilon^j\gamma_\mu$, one obtains
\begin{eqnarray}
  i_VF^I&=&-D(fX^I)\label{diff 7}\\
  -\left(\frac{1}{4}Q_{IJ}-\frac{3}{8}X_IX_J\right)F^J_{\mu\nu}
  (\star J^i)_{\rho}^{\ \ \mu\nu}&=&-\frac{3}{4}(J^i)_\rho^{\ \ \mu}
  D_\mu X_I\ ,\label{diff 8}
\end{eqnarray}
where $D$ without subscript denotes exterior $K$-gauge covariant
derivatives. Finally, contracting it with
$\bar\epsilon^j\gamma_{\mu\nu}$, one obtains equations similar to
those in \cite{gu-sa}. We do not record them as they will not be
used in this paper.

\end{document}